\documentclass{csmadapter}

\jvol{XX}
\jnum{XX}
\paper{X}
\jmonth{August}
\jname{IEEE CONTROL SYSTEMS}
\pubyear{2024}
\usepackage{bm}
\usepackage{optidef}
\usepackage{hyperref}

\usepackage[colorinlistoftodos]{todonotes}
\usepackage{multirow}
\usepackage{pifont}
\usepackage{algorithm} 
\usepackage{algpseudocode} 
\usepackage{physics}
\usepackage[normalem]{ulem}
\newcommand{\cmark}{\textcolor{green}{\ding{51}}}%
\newcommand{\xmark}{\textcolor{red}{\ding{55}}}%

\usepackage{xspace}
\newcommand{\acronym}[1]{{#1}\xspace}

\newcommand{\Bv}{\mathbf v}
\newcommand{\Bu}{\mathbf u}

\newcommand{\RR}{\mathbb R}

\newcommand{\sdbarfigfullwidth}[2]{%
    \end{multicols}
    \let\thefigure\thesfigure
    \renewcommand{\thesfigure}{S\arabic{sfigure}}%
    \sdbarfig{#1}{#2}
    \begin{multicols}{2}
}

\begin{document}

\title{Traffic Control via Connected and Automated Vehicles\stitle{An Open-Road Field Experiment with 100 CAVs}}

\author{
    JONATHAN W.~LEE\textsuperscript{*,1},
    HAN WANG\textsuperscript{*,1},
    KATHY JANG\textsuperscript{*,1},
    AMAURY HAYAT\textsuperscript{\textdagger,1},
    MATTHEW BUNTING\textsuperscript{\textdaggerdbl,1},
    
    ARWA ALANQARY\textsuperscript{*,2},
    WILLIAM BARBOUR\textsuperscript{\textdaggerdbl,2},
    ZHE FU\textsuperscript{*,2},
    XIAOQIAN GONG\textsuperscript{\textsection,2},
    GEORGE GUNTER\textsuperscript{\textdaggerdbl,2},
    SHARON HORNSTEIN\textsuperscript{\textbardbl,2},
    ABDUL RAHMAN KREIDIEH\textsuperscript{*,2},
    NATHAN LICHTLÉ\textsuperscript{*,\textdagger,2},
    MATTHEW W.~NICE\textsuperscript{\textdaggerdbl,2},
    WILLIAM A.~RICHARDSON\textsuperscript{\textdaggerdbl,2},
    ADIT SHAH\textsuperscript{*,2},
    EUGENE VINITSKY\textsuperscript{*,2},
    FANGYU WU\textsuperscript{*,2},
    SHENGQUAN XIANG\textsuperscript{\textbardbl\textbardbl,2},
    
    SULAIMAN ALMATRUDI\textsuperscript{*,3},
    FAHD ALTHUKAIR\textsuperscript{*,3},
    RAHUL BHADANI\textsuperscript{\textdaggerdbl,3},
    JOY CARPIO\textsuperscript{*,3},
    RAPHAEL CHEKROUN\textsuperscript{*,3},
    ERIC CHENG\textsuperscript{*,3},
    MARIA TERESA CHIRI\textsuperscript{\P,3},
    FANG-CHIEH CHOU\textsuperscript{*,\textdagger\textdagger,3},
    RYAN DELORENZO\textsuperscript{\textdaggerdbl\textdaggerdbl,3},
    MARSALIS GIBSON\textsuperscript{*,3},
    DEREK GLOUDEMANS\textsuperscript{\textdaggerdbl,3},
    ANISH GOLLAKOTA\textsuperscript{*,3},
    JUNYI JI\textsuperscript{\textdaggerdbl,3},
    ALEXANDER KEIMER\textsuperscript{*,3},
    NOUR KHOUDARI\textsuperscript{**,3},
    MALAIKA MAHMOOD\textsuperscript{\textdaggerdbl\textdaggerdbl,3},
    MIKAIL MAHMOOD\textsuperscript{\textdaggerdbl\textdaggerdbl,3},
    HOSSEIN NICK ZINAT MATIN\textsuperscript{*,3},
    SEAN MCQUADE\textsuperscript{\textdaggerdbl\textdaggerdbl,3},
    RABIE RAMADAN\textsuperscript{**,3},
    DANIEL URIELI\textsuperscript{\textbardbl,3},
    XIA WANG\textsuperscript{\textdaggerdbl,3},
    YANBING WANG\textsuperscript{\textdaggerdbl,3},
    RITA XU\textsuperscript{*,3},
    MENGSHA YAO\textsuperscript{**,3},
    YILING YOU\textsuperscript{*,3},
    GERGELY ZACH\'{A}R\textsuperscript{\textdaggerdbl,3},
    YIBO ZHAO\textsuperscript{*,3},
    
    MOSTAFA AMELI\textsuperscript{\P\P, *,4}
    MIRZA NAJAMUDDIN BAIG\textsuperscript{\textdagger\textdagger,4},
    SARAH BHASKARAN\textsuperscript{*,4},
    KENNETH BUTTS\textsuperscript{\textsection\textsection,4},
    MANASI GOWDA\textsuperscript{*,4},
    CAROLINE JANSSEN\textsuperscript{\textdaggerdbl,4},
    JOHN LEE\textsuperscript{*,4},
    LIAM PEDERSEN\textsuperscript{\textdagger\textdagger,4},
    RILEY WAGNER\textsuperscript{\textdaggerdbl,4},
    ZIMO ZHANG\textsuperscript{*,4},
    CHANG ZHOU\textsuperscript{*,4},
    
    DANIEL B.~WORK\textsuperscript{\textdaggerdbl},
    BENJAMIN SEIBOLD\textsuperscript{**},
    JONATHAN SPRINKLE\textsuperscript{\textdaggerdbl},
    BENEDETTO PICCOLI\textsuperscript{\textdaggerdbl\textdaggerdbl},
    {MARIA LAURA} {DELLE MONACHE}\textsuperscript{*}, AND
    ALEXANDRE M.~BAYEN\textsuperscript{*}
}
\affil{
    *--University of California, Berkeley\\
    \textsuperscript{\textdagger}--\'Ecole des Ponts Paristech, Marne la Vallée\\
    \textsuperscript{\textdaggerdbl}--Vanderbilt University\\
    \textsuperscript{\textsection}--Arizona State University\\
    \textsuperscript{\textbardbl}--General Motors\\
    \textsuperscript{\P\P}--Université Gustave Eiffel, COSYS-GRETTIA\\
    \textsuperscript{\textbardbl\textbardbl}-- Peking University\\
    \textsuperscript{\P}--Queen's University\\
    \textsuperscript{**}--Temple University\\
    \textsuperscript{\textdagger\textdagger}--Nissan North America\\
    \textsuperscript{\textdaggerdbl\textdaggerdbl}--Rutgers University-Camden\\
    \textsuperscript{\textsection\textsection}--Toyota North America\\
    \textsuperscript{1}--These authors contributed equally\\
    \textsuperscript{2}--These authors contributed equally\\
    \textsuperscript{3}--These authors contributed equally\\
    \textsuperscript{4}--These authors contributed equally
}

\maketitle

\dois{}{}

\newpage
\begin{summary}
\summaryinitial{T}he CIRCLES project aims to reduce instabilities in traffic flow, which are naturally occurring phenomena due to human driving behavior. Also called ``phantom jams" or ``stop-and-go waves,'' these instabilities are a significant source of wasted energy~\cite{FlynnKasimovNaveRosalesSeibold2009, giammarino2020traffic}. Toward this goal, the CIRCLES project designed a control system referred to as the MegaController by the CIRCLES team, that could be deployed in real traffic. Our field experiment leveraged a heterogeneous fleet of 100 longitudinally-controlled vehicles as Lagrangian traffic actuators, each of which ran a controller with the architecture described in this paper. The MegaController is a hierarchical control architecture, which consists of two main layers. The upper layer is called Speed Planner, and is a centralized optimal control algorithm. It assigns speed targets to the vehicles, conveyed through the LTE cellular network. The lower layer is a control layer, running on each vehicle. It performs local actuation by overriding the stock adaptive cruise controller, using the stock on-board sensors.
The Speed Planner ingests live data feeds provided by third parties, as well as data from our own control vehicles, and uses both to perform the speed assignment. The architecture of the speed planner allows for modular use of standard control techniques, such as optimal control, model predictive control, kernel methods and others. The architecture of the local controller allows for flexible implementation of local controllers. Corresponding techniques include deep reinforcement learning, model predictive control and explicit controllers. Depending on the vehicle architecture, all onboard sensing data can be accessed by the local controllers, or only some. 
Likewise, control inputs vary across different automakers, with inputs ranging from torque or acceleration requests for some cars, and electronic selection of ACC set points in others.
The proposed architecture technically allows for the combination of all possible settings proposed above, that is $\{$Speed planner algorithms$\}\times\{$local controller algorithms$\}\times\{$full or partial sensing$\}\times\{$torque or speed control$\}$. Most configurations were tested throughout the ramp up to the MegaVandertest. 
\end{summary}

\chapterinitial{T}he promise of generating societal benefit from autonomous vehicle technology has long been capturing the imagination of researchers, legislators, and popular culture. Its footprints can be seen in the earliest stages of modern AV research and development in the United States. 
AV technology advancements,
spanning over three decades of work, are sometimes broken down into three generations:

\noindent\textbf{Generation One.} In the 1990's, the \acronym{Federal Highway Administration} (FHWA) established the \acronym{National Automated Highway System Consortium} (NAHSC) to demonstrate the potential of an automated vehicle and highway system for societal benefit~\cite{stevens1996automated}. The consortium partners included University of California, Berkeley, \acronym{California Partners for Advanced Transit and Highways} (PATH), and \acronym{General Motors} (GM), among several others. In the implementation of Demo '97 on Interstate 15 in San Diego~\cite{tan1998demonstration}, California, the consortium introduced and/or brought to realization concepts such as \acronym{Adaptive Cruise Control} (ACC), \acronym{Vehicle-to-Vehicle} (V2V) Communication, and \acronym{Cooperative ACC} (CACC)~\cite{shladover2015cooperative,swaroop1994comparision,ioannou1993autonomous}. The demonstrations involved high-speed platooning, made possible by V2V and CACC, as well as infrastructure sensor integration. Moreover, it laid the foundation for modern \acronym{automated highway system} AHS architectures, emphasizing the importance of vehicle-to-vehicle (V2V) and vehicle-to-infrastructure (V2I) communications~\cite{girault2004hybrid}. Some of the key challenges and barriers faced at the time included scalability due to infrastructure, reliable communication systems, and accurate vehicle positioning. Its eventual demise came in the late 1990's with tightening research budgets at USDOT.

\noindent\textbf{Generation Two.} In the 2000's, 
\acronym{the Defense Advanced Research Projects Agency} (DARPA) 
introduced the DARPA Grand Challenge. In the first Grand Challenge (2004), a \$1 million prize was offered to any team whose autonomous vehicle could complete a 150-mile course in the Mojave Desert; no teams were successful. The second Grand Challenge (2005) doubled the prize money, and five teams were successful in completing the course~\cite{buehler20072005}. The third Grand Challenge (2007) moved to an urban environment. 
The DARPA Grand Challenge competitions were notable for creating a renewed interest in autonomous vehicles and directly supported efforts to commercialize successful technologies~\cite{thrun2006stanley} in the following decade.

\noindent\textbf{Generation Three.} In the 2010's, many major automotive manufacturers (such as GM, BMW, audi, and Tesla) and technology companies (such as Google/Waymo, Uber, Lyft) began developing autonomous vehicle technology for commercialization purposes. The investments have generated an industrial ecosystem centered on hardware, software, and services designed to advance fully autonomous driving. Today multiple companies are testing 
AVs
in numerous public roads in the US, with Waymo most notably operating an autonomous ride-share service in Chandler, AZ without a safety driver in the vehicle, and Cruise, which operates autonomous shuttles in San Francisco. At the same time, Level 1 and Level 2 automated vehicle technologies, including ACC, are now widely available in the commercial market. These commercial systems predominantly focus on safety and driver comfort~\cite{xiao2010comprehensive}, rather than improved transportation system efficiency~\cite{gunter2020commercially}.

Our work brings full circle the nascent imaginings from the early 1990's. The Generation One promise of automating highways for societal benefit began to manifest in the late 2010's within our team and in several groups around the world. In 2016, a group led by co-authors Delle Monache, Piccoli, Seibold, Sprinkle, and Work replicated the Sugiyama experiment~\cite{Sugiyama_2008} of string instability in a ring road; in what is known as the ``Arizona ring experiment,'' they further demonstrated that phantom jams can be reduced using partially \acronym{automated vehicle} (AV) technologies and specially-designed algorithms~\cite{stern2018dissipation}. This work was followed shortly by co-author Bayen's lab at UC Berkeley with the development of \texttt{Flow}, a software package to interface then-state-of-the-art microsimulation software with \acronym{deep reinforcement learning} (deep-RL) libraries~\cite{flow, wu2017emergent, vinitsky2018benchmarks, jang2019simulation,AAMAS21-cui, zhang2023learning}. With \texttt{Flow}, the UC Berkeley team was able to independently train an AV controller to replicate the findings of the Arizona ring experiment~\cite{wu2018stabilizing} and generalize them to a variety of other settings, such as freeway and urban traffic~\cite{pmlr-v87-vinitsky18a}. (Note the distinction in terminology between ``autonomous vehicle'' and ``automated vehicle,'' specifically because our past and present work focuses on automated longitudinal control as Lagrangian traffic controllers.)


In 2019, co-authors Bayen, Piccoli, Seibold, Sprinkle, and Work united to form the \acronym{Congestion Impacts Reduction via CAV-in-the-loop Lagrangian Energy Smoothing} (CIRCLES) Consortium (see \url{https://circles-consortium.github.io/})~\cite{CIRCLES_AMR2020, CIRCLES_AMR2021, CIRCLES_AMR2022}, joined shortly thereafter by co-authors Lee and Delle Monache, each leading various aspects of the CIRCLES project. The CIRCLES project seeks to extend these prior research efforts to real traffic. To achieve our goals, our group designed a modular, hierarchical control framework, consisting of a centralized \acronym{Speed Planner} and decentralized \acronym{Vehicle Controllers}, and implemented it on 100 vehicles in a large-scale field operational test, dubbed the \acronym{MegaVanderTest}~(MVT). 


This article presents our control system design and subsequent analysis of field test results.
Diverse candidates for each module of the framework are developed utilizing cross-disciplinary knowledge and tools, including ordinary differential equation/partial differential equation (ODE/PDE)-based flow control~\cite{WangSpeedplanner},
deep-RL~\cite{JangReinforcement}, stabilization theory \cite{bacciotti2005liapunov}, 
functional analysis
\cite{yosida2012functional}, 
optimal control on microscopic and macroscopic systems \cite{bressan2007introduction,casas2017optimal}, 
approximation theory \cite{trefethen2019approximation}, 
mean-field limits \cite{golse2016dynamics}, 
non-entropic solutions to hyperbolic systems \cite{baiti2001uniqueness}, \acronym{model predictive control} (MPC) via  
\acronym{linearly constrained quadratic programming} (LCQP) \cite{wu2023}, 
kernel smoothing \cite{fu2023cooperative}, 
traffic flow theory \cite{garavello2016models}, 
variable speed limit \cite{lu2014review}, 
and many other areas. 
The control system was then evaluated and tested in the open road as part of the MVT, the largest deployment of AVs designed to smooth traffic flow. In this test, we deployed 100 AVs on \acronym{Interstate 24} (I-24) near Nashville, TN in November 2022. The experiment coincided with the debut of I-24 MOTION (\url{https://i24motion.org}) \cite{gloudemans202324,gloudemans2020interstate}, a four-mile section of I-24 near Nashville, TN to capture ultra-high resolution trajectory data of all vehicles. 


\begin{pullquote}
This work presents a modular, hierarchical control framework, consisting of a centralized Speed Planner and decentralized Vehicle Controllers, as it was used during the MegaVanderTest. 
\end{pullquote}

\section{Design of the Controller Architecture}
\begin{figure*}[ht]
\centerline{\includegraphics[width=38.0pc]{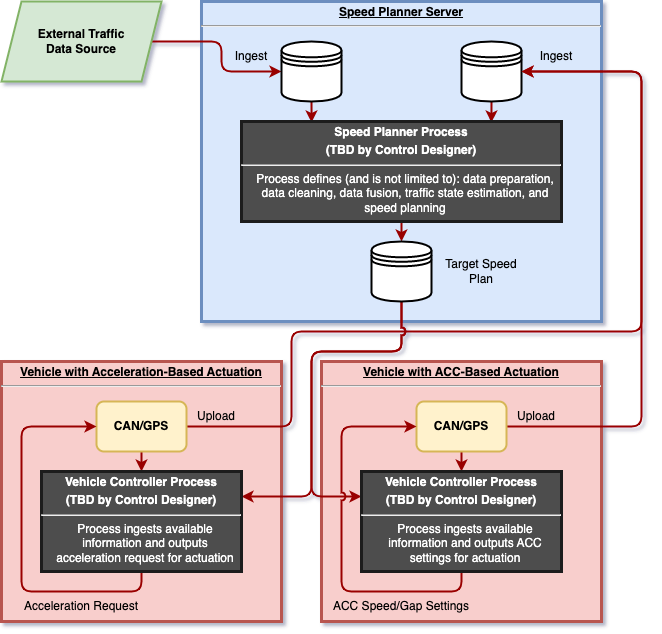}}
\caption{Architectural framework of the MegaController. The hierarchical and modular nature of the design allows for greater flexibility in design decisions and dealing with varied sensing and actuation capabilities of the heterogeneous fleet. The blue box represents the centralized Speed Planner unit, and the red boxes represent decentralized Vehicle Controllers, which are vehicle-dependent (that is, each vehicle has a different control architecture and thus requires a different control paradigm). The components work in concert to achieve higher level goals of flow smoothing.\label{fig:framework}}
\end{figure*}

We introduce the \textbf{MegaController}, a control framework, depicted in Figure \ref{fig:framework}, for the mixed autonomy traffic flow problem. Mixed autonomy refers to the setting in which some vehicles are  automated, while other vehicles are manually controlled (level 0 automation). The primary design paradigm of the proposed control framework is to achieve two goals: hierarchy for task allocation and modularity for control flexibility.

We designed a hierarchical structure to efficiently coordinate the control goals between macroscopic traffic flow optimization~\cite{bayen2022control} and microscopic vehicle control \cite{alanqary2023optimal,SternETAL2016,bhadani449} and to efficiently solve the computational task allocation problem between the server and vehicle sides. There is an inherent interplay between the two components, as the Speed Planner informs the Vehicle Controller of downstream events, and the vehicles are instrumented to report observations to the server for data aggregation.

\subsection{Use of modular design}
A modular structure is used for two important purposes: (1) to facilitate a diversity of controller design approaches, and (2) to enable a heterogeneous vehicle fleet with different sensing and actuation capabilities. 


As long as each controller utilizes the available interfaces, it is technically straightforward to exchange different controllers when using a modular design. This decision allows us to support a wide range of expertise from our team who may explore many different designs, each with a different technical focus or approach. While there may be subtleties for stability, convergence, and other properties if integration decisions are made strictly through type matching, our modular approach mitigates a key challenge: the dynamics of the roadway and our ability to control it may not be fully understood until testing begins. Since testing may only be possible at a scale when we can evaluate its impact, we will be very close to the full test deadline when we make final decisions. By embracing a modular approach, we can use data from tests that concluded less than 24 hours earlier to decide what controller to run during the following test.

The fleet of vehicles used in the MVT represented 3 different years, makes, and models. The fleet was composed of 2023 Nissan Rogue, 2022 Cadillac XT5, and 2020-2021 Toyota Rav4. 
Trade secrets and design differences between the systems on these vehicles mean that it is not possible to have a uniform system interface for sensing and control. 
For example, although each car we used has adaptive cruise control which uses forward facing radar for safety, the data from those sensors were not available to our system for all vehicle types. 
Designing
a controller that can work on only one vehicle type may give greater controllability, but reduces the potential impact of the design when deployed at a societal scale. 

Thus it is a design strength that our system can operate across vehicles with myriad different sensing and control modalities. 
While the interfaces to the Speed Planner remain the same, different vehicles will have different implementations that have completely unique characteristics compared to other vehicles, similar to textbook design patterns in software engineering for producer/consumer architectures. This allows a flexibility to adapt to a heterogeneous fleet, theoretically opening the door to many other vehicle configurations and constraints.

In the MegaController design, there are two distinct components:
\begin{itemize}
    \item~A server-side \textbf{Speed Planner}, which is a centralized planner unit that provides high-level macroscopic speed suggestions based on periodic state updates from the distributed vehicles and external macroscopic data source, like INRIX employed in this study. The algorithms deployed on the server-side are designed to handle computationally-heavy data aggregation and macro-state tasks.
    \item~A vehicle-side \textbf{Vehicle Controller}, which is a networked decentralized controller \cite{baillieul2007} 
    that commands local actuation of the vehicle. The algorithms deployed here take into consideration the target
    speed
    suggested by the server-side planner unit, the latency of that information, and any observations from the vehicle's onboard sensors.
\end{itemize}
Borrowing conceptually from object-oriented programming, the interfaces are agreed upon \emph{a priori}, allowing an abstraction of individual components for ease of design and testing.


\begin{pullquote}
The design paradigm of the control framework achieves two goals: hierarchy for task allocation, and modularity for control flexibility.
\end{pullquote}


\subsection{Speed Planner}

The objective of the Speed Planner is to optimize the overall traffic flow efficiency by providing macroscopic guiding speeds for AV fleets using realtime traffic data from multiple heterogeneous sources. While optimality for our deployment was defined by flow efficiency, 
our
framework allows for arbitrary objective functions.

The
data sources at our disposal are 
the set of recent 
vehicle ping messages from our AV fleet using our custom hardware and software stack (detailed later in Figure~\ref{fig:caninjection}),
as well as vehicular velocity data within distinct sections of highway through INRIX, a provider of live data for traffic, parking, and other transportation data. INRIX data has been studied in the literature for its applicability in traffic estimation, and validated by researchers for its accuracy using \textit{post hoc} methods \cite{inrix2019}. 
Utilizing these heterogeneous data sources presented some key challenges and considerations, namely unknown latency of external data sources and differences in spatial and temporal resolution of those data sources when fused with our 
vehicle message data.
Although the specific aggregation algorithm is not publicly known, INRIX data latency is explored in \cite{KIM201459}, which describes delays between 3-5 minutes, and in some cases up to 10 minutes. A summary of 
data
%
specifications can be found in Table~\ref{table:data_sources}.

\begin{table}[htbp]
\begin{center}
\begin{tabular}{l||l|l}
Data Source:           & INRIX             & AV Ping    \\ \hline\hline
Period (seconds)    & 60                & 1          \\
Latency (seconds)      & $\sim$180         & negligible \\
Segment Length (miles) & $\sim$0.5 (varied) & N/A       \\
Lateral aggregation & Lane averaged & Lane specific

\end{tabular}
\caption{A summary table of the data sources for traffic state estimation (and subsequent processes). INRIX aggregates data from a fleet of vehicles on the road.
The AV Ping data come from our own fleet, where each car posts speeds to the server through an API that contains speed (from CAN) and positioning (from GPS), as well as timing and vehicle identity information.}
\label{table:data_sources}
\end{center}
\end{table}

Targeting the challenges, the two main functions of the Speed Planner are (1) \acronym{Traffic State Estimation} (TSE) enhancement and (2) the design of the Speed Plan, an ideal target speed profile. The TSE enhancement module is designed to eliminate the effects of inherent latency in the third-party data source and to improve the spatial and temporal resolution of the input traffic data. The target design module generates Speed Plans with the goal of reducing vehicle energy consumption and increasing the overall throughput of the traffic flow. Various approaches are utilized to develop the target design module, including kernel smoothing~\cite{fu2023cooperative}, ODE/PDE-based flow control~\cite{WangSpeedplanner} and vehicle trajectory optimization~\cite{alanqary2023optimal}.



\subsection{Vehicle Controller}

The Vehicle Controller enables each car to use traffic information provided by the Speed Planner in context with local state, such as the distance to the vehicle in front, relative speed, etc. The Vehicle Controller aims to improve the flow of traffic while ensuring the safety of both the AV and other vehicles around it.

Input data for the Vehicle Controller are obtained from the in-vehicle network, and may include other inputs such as the target speeds from the Speed Planner. 
A subset of data from the CAN bus are made available to the vehicle controller, and vary depending on the vehicle make/model/year. Realizing control algorithms on vehicles required new computer hardware, electronics, and software development ~\cite{bunting2021libpanda,elmadani2021can,lichtle2022deploying,bhadani2023approaches,nice2023middleware,nice2023parameter} because of the heterogeneity in vehicle data and control interfaces. 

Depending on the vehicle platform, we have explored options including speed-based control \cite{bunting2022data,chou2021reachability}, acceleration-based control~\cite{lichtle2022deploying, lichtle2023traffic}, and ACC-based control~\cite{JangReinforcement}. 
Speed-based control was initially explored, but then deprecated as acceleration based control best matched desired goals from algorithm designers. The acceleration-based and ACC-based approaches fit into the overall architectural design. Torque-based control (that is, applying the requisite torque directly to the gears) would be the most prescriptive control with tightest actuation, but it also requires a level of CAN bus access that prevents quick return of the vehicle to stock state. The majority of the vehicles used in the experiment were stock vehicles that were later returned to the manufacturer, so all changes to enable experimental control needed to be minimal and non-invasive changes to the vehicle. Therefore, for this deployment, our architecture included two types of control capabilities on our heterogeneous fleet, which are summarized in Table~\ref{table:vehicle_capabilities}.



\begin{table}[htbp]
\begin{center}
\begin{tabular}{cl||c|c}
\multicolumn{2}{l||}{Vehicle(s):} & Toyota RAV4 & \shortstack{Nissan Rogue,\\Cadillac XT5} \\ 
\hline
\multicolumn{2}{l||}{Actuation Type:} & Acceleration & ACC \\ 
\hline\hline
\multicolumn{1}{l|}{\multirow{6}{*}{\rotatebox{90}{Sensing}}} & Ego Position & \cmark & \cmark \\
\multicolumn{1}{l|}{} & Ego Speed & \cmark & \cmark \\
\multicolumn{1}{l|}{} & Ego ACC Speed Setting & \cmark & \cmark \\
\multicolumn{1}{l|}{} & Ego ACC Gap Setting & \cmark & \cmark \\
\multicolumn{1}{l|}{} & Leader Space Gap & \cmark & \xmark \\
\multicolumn{1}{l|}{} & Leader Minicar & \cmark & \cmark \\
\multicolumn{1}{l|}{} & Leader Relative Speed & \cmark & \xmark 
\end{tabular}
\end{center}
\caption{A summary table of the sensing and actuation available to our system, for the heterogeneous vehicle fleet. Space gap refers to the distance between the ego vehicle's front bumper and the leader vehicle's rear bumper. The minicar refers to a boolean indication whether the ego vehicle's sensors detect the presence of a leader vehicle (an approximate 80-100 meter maximum distance). Relative speed is the leader vehicle's speed minus the ego vehicle's speed.}
\label{table:vehicle_capabilities}
\end{table}


\begin{sidebar}{Adaptive Cruise Control Modeling}
\noindent
by Fang-Chieh Chou, Jonathan W.~Lee
\section[Adaptive Cruise Control Modeling]{}\label{sbar-ACC}
\vspace{-5mm}

\setcounter{sequation}{0}
\renewcommand{\thesequation}{S\arabic{sequation}}
\setcounter{stable}{0}
\renewcommand{\thestable}{S\arabic{stable}}
\setcounter{sfigure}{0}
\renewcommand{\thesfigure}{S\arabic{sfigure}}

\sdbarinitial{A}\emph{daptive Cruise Control} (ACC) is an advanced driver assistance system that automatically operates vehicles at a set driving speed or keeps a set driving gap with respect to a leading vehicle. 
Operation of ACC requires a speed setting and a gap setting input by the driver. When there is no leading vehicle, ACC automatically adjusts the vehicle's driving speed to the speed setting preset by the driver. When there is a leading vehicle ahead, the system automatically regulates the gap with respect to the leading vehicle to a separation set by the driver, while not driving faster than the speed setting.

An ACC-controlled car therefore be modeled as a dynamical system composed of two control modes: speed-control mode and gap-control mode. The system switches between two modes depending on the proximity and speed of a leading vehicle. When there is no nearby leading vehicle, the system is in speed-control mode, in which the system tracks the speed setting. On the other hand, when the system is in gap-control mode, the system tracks both the leading vehicle speed and the gap setting. 

To model an ACC-controlled car, we used parametric models for each control mode.
Mathematically, the speed-control mode can be written as
\begin{equation}
    a_e = f_p(v_e , v_\text{ref})\;,
\end{equation}
where $a_e$ is the acceleration of the ego vehicle, $v_e$ is the speed of the ego vehicle, and $v_\text{ref}$ is the speed setting. $f_p: \mathbb{R}\times\mathbb{R} \rightarrow \mathbb{R}$ is a parametric model of parameters $p$. 

Gap-control mode can be written as
\begin{equation}
    a_e = f_q(v_e , v_l , d_{el}, g_\text{ref})\;,
\end{equation}
where $v_l$ is speed of leading vehicle, $d_{el}$ is the space gap between the ego vehicle and the leading vehicle, and $g_\text{ref}$ is the gap setting. $f_q : \mathbb{R}\times\mathbb{R}\times\mathbb{R}\times\mathbb{R} \rightarrow \mathbb{R}$ is a parametric model of parameters $q$. 

To fit parameters $p$ and $q$ for speed-control mode and gap-control mode, field experiments are carried out using a Nissan Leaf vehicle equipped for data collection. For the speed-control mode, data collection is done with a variety of initial speeds and speed settings so that dynamic responses with different speed changes can be collected. For the gap-control mode, highway driving data under different traffic conditions is collected so that dynamic response over a range of speeds can be covered in the data set. The collected data is firstly smoothed to reduce sensor noise. Some outliers of abnormal driving conditions (for example, temporally near cut-in and cut-out events) are removed before fitting the models. The model is validated by comparing the dynamic response of the model in simulation to the response of the real system in field experiments. While the models are fitted for the Nissan Leaf, they are assumed to be acceptable approximations for the ACC systems of other vehicles in our AV fleet.
\end{sidebar}


\subsubsection{Acceleration-based Controller}
While acceleration controllers don't have natural safety enforcement such as the one speed controllers have, we have found a way to add safety enforcements.
Using new approaches in \cite{gunter2022experimental}, we demonstrated that it was possible to constrain unsafe accelerations when they were passed to the vehicle. This opened the door to use of acceleration-based control on cars that support it.  

Acceleration-based control is a natural analog to how most of us drive: we press down on the accelerator pedal when we want to go faster, and we either let off the accelerator, or press the brake pedal, when we want to go slower.

When making control requests to the car, we provide a desired input over the vehicle's CAN, which is interpreted by the powertrain and ACC subsystems of the car. This desired input is not strictly a direct acceleration command: it is processed by the vehicle's ACC system 
which interprets the commanded acceleration and decides whether it will meet it (a) by actuating the brake, (b) by decreasing acceleration through reducing the throttle, (c) by keeping the throttle constant, or (d) by increasing the throttle. Examples of how these dynamics from step inputs have been observed can be found in \cite{bunting2022data}.
The specific input selected by the car is a function of many different vehicle dynamics and perhaps even trade secrets for reducing wear-and-tear of parts, so there is no single mapping or predictive function that easily codifies the transfer function of the system once an acceleration is requested. 


Using acceleration-based control \textit{implicitly} adds safety challenges, as the controller must ignore all existing acceleration commands from the car---including any acceleration commands that come from the safety subsystems on board. Safety wrappers such as \cite{gunter2022experimental} are needed to prevent rear-end collisions, and additional care must be taken to evaluate external sensors. Further, environmental challenges such as lane-changes by other vehicles require our system to ensure that our dynamics do not inadvertently cause their own stop-and-go waves. We decompose the behavior of this controller into two modes:


\begin{itemize}
    \item ~A \textbf{base controller} prescribes a performance-based desired acceleration under ``ideal'' conditions.  The base controller is designed to output an acceleration that satisfies a specific performance-based goal, for example, reducing energy footprint or minimizing the magnitude of accelerations. In general, care is taken during its design-phase to ensure that other aspects of the traffic flow are not negatively impacted. Note that the base controller should still, by design, be safe and collision-free, under the assumption that accelerations are actuated perfectly.
    \item ~A \textbf{lane-change recovery controller} modulates the base controller's desired acceleration in the event of discontinuities in the lead-vehicle space gap. The acceleration output of the base controller can change dramatically if there is a time-discontinuity in the input signals. Most commonly, this appears in the form of a lane-change by another vehicle in front of the ego vehicle. Since a lane-change is out of the ego vehicle's control, it can theoretically put the ego vehicle into an unsafe state if the objective were to control to a time gap with discontinuous input. This mode recognizes that discrete changes in the space gap are not controller failures, so the objective of this controller is to comfortably return the vehicle to a safe state where the base controller can resume actuation.
\end{itemize}


\subsubsection{ACC-based Controller}
Another option for controlling vehicles is to update vehicle's \emph{Adaptive Cruise Control} (ACC) set points electronically. This mimics what the driver can do through buttons on the steering wheel, and has the benefit of keeping the safety features of the car's ACC systems in the loop. 
This approach is contrasted with the acceleration-based approach in a few ways, notably:
\begin{itemize}
    \item The controllability of the system is affected and the bandwidth of the controller is likely reduced, since the stock ACC has its own gains and modes.
    \item The multi-model design features of the acceleration-based approach (namely the lane-change recovery controller and safety wrapper) are not needed here, as those features are built into the stock ACC algorithm.
\end{itemize}
To understand the changes in controllability, we explored the vehicle's ACC dynamics with a goal to make design decisions on whether rate limits or other constraints on the input signals would be required for stability or performance reasons.
Every vehicle make and model equipped with ACC has its own unique mapping of state space (for example, ego speed, space gap, leader speed, etc.) and ACC inputs (speed setting and gap setting) to actuated acceleration. Architecturally, we take as given that this model is available for the design of the base ACC controller. For a more detailed discussion of ACC and our specific approach to obtaining a model for our algorithm design, see ``\nameref{sbar-ACC}".

As a result, the only component of the ACC-based controller that requires specific designing is the base controller. Just as with the acceleration-based controller, the base controller is performance-based, striving for some performance goal. Similar design philosophies are applied as before. Uniquely different, as noted above, is that this base controller does not need to explicitly consider safety and vehicle-specific dynamics---the controller design can implicitly account for the vehicle-specific dynamics by utilizing the ACC model in a feedback loop.


\begin{pullquote}
The MegaVanderTest deployed 100 vehicles in November of 2022, and was the largest coordinated open-road test to smooth traffic flow. 
\end{pullquote}

\section{Design of Controller Components}
Here, we describe in detail how each of the modular components was designed. The discussion is in a parallel structure to the previous section, with particular focus on controllers that were actually implemented for deployment. In the lifespan of the project, we explored many approaches for designing the controller. See ``\nameref{sbar-MPC}'' and ``\nameref{sbar-OC}'' for some notable research and designs that were not deployed but contributed key concepts that helped the ultimate implementation.

\begin{sidebar}{Macroscopic ODE/PDE models}
\noindent
by Xiaoqian Gong
\section[Macroscopic ODE/PDE models]{}\label{sbar-MOPM}
\vspace{-5mm}
\renewcommand{\thesequation}{S\arabic{sequation}}
\renewcommand{\thestable}{S\arabic{stable}}
\renewcommand{\thesfigure}{S\arabic{sfigure}}

\sdbarinitial{H}ere we model the traffic dynamics in the presence of $M$ AVs and $N$ human-driven vehicles on a single lane from the microscopic perspective using systems of ordinary differential equations. We assume that only AVs can be controlled and have a greater impact on the vehicle population than human-driven vehicles. Let $T>0$ be a fixed time horizon, $I_M = \{1, \cdots, M\}$ and $I_N = \{1, \cdots, N\}$ the index sets of AVs and human-driving vehicles, respectively. Denote by $(x,v) \in \mathbb{R}^N\times \mathbb{R}^N_{\geq 0}$ and $(y,w)\in \mathbb{R}^M\times \mathbb{R}^M_{\geq 0}$ the position-velocity vectors of human-driven vehicles and AVs, respectively. To represent the positions and velocities of the $M$ AVs during the time interval $[0, T]$, we define the time-dependent atomic probability measure on $\mathbb{R} \times \mathbb{R}_{\geq 0}$,
also referred to as the \emph{empirical measure}, as
\begin{align}
\label{atomic_measure_av}
	\mu_{M} (t)& = \frac{1}{M} \sum \limits_{i=1}^{M} \delta_{\left(y_i(t), w_i(t)\right)}\;, \quad t\in [0,T] .
\end{align}
Alternatively, we can represent solutions as a measure
supported on absolutely continuous trajectories $t \in [0,T] \mapsto (y_i(t), w_i(t)) \in \mathbb{R} \times \mathbb{R}_{\geq 0}, i \in I_M$. Similarly, we use the atomic measure 
\begin{align}
\label{eqn: mu_N}
\mu_{N} (t)& = \frac{1}{N} \sum \limits_{j=1}^{N} \delta_{\left(x_j(t), v_j(t)\right)}\;, \quad t \in [0,T]
\end{align}
to track the positions and velocities of the $N$ human-driven vehicles during the time interval $[0,T]$. 
The dynamics of the $M+N$ vehicles is given as follows: 
\begin{equation}\label{ODEs}
\begin{aligned}
\dot{y}_i &= w_i\;,\\
\dot{w}_i &= \left(H_1*_1(\mu_{N} + \mu_M)+H_2*(\mu_{N}+\mu_{M})\right)(y_i, w_i) + u_i\;, \, i \in I_M\;, \\
\dot{x}_j &= v_j\;,\\
\dot{v}_j &= \left(H_1*_1(\mu_{N} + \mu_M)+H_2*(\mu_{N}+\mu_{M})\right)(x_j, v_j)\;, \, j \in I_N\;,
\end{aligned}
\end{equation}
where the convolution kernels $H_1 \colon \mathbb{R} \times \mathbb{R}_{\geq 0} \to \mathbb{R}$ 
and $H_2 \colon \mathbb{R} \times \mathbb{R} \to \mathbb{R}$ 
represents a microscopic model, such as Optimal Velocity \cite{bando1995dynamical}, Follow-the-Leader \cite{gazis1961nonlinear} or
a combination of them \cite{Hayat2023theory}. Here $*_1$ is the convolution concerning the first variable, and $u_i \colon [0, T] \to \mathbb{R}$ are measurable controls for $i \in I_M$ influencing the time-evolution of AVs.
Given initial data $(x(0), v(0), y(0), w(0)) = (x_0, v_0, y_0, w_0) \in \mathbb{R}^N \times \mathbb{R}_{\geq 0}^N \times \mathbb{R}^M \times \mathbb{R}_{\geq 0}^M$, the existence and uniqueness of solutions to system (\ref{ODEs}) can be proved using Carath\'eodory theorem. This is a consequence of the fact that the two convolution kernels $H_1$ are locally Lipschitz with sub-linear growth.
For more detailed discussions, we refer the readers to~\cite{PR13}.

Now we consider modeling mixed traffic dynamics when the number of human-driven vehicles is much greater than the number of AVs. This allows us to pass to the mean-field limit of the system (\ref{ODEs}) with the number of human-driven vehicles formally going to infinity, that is $N \to \infty$. The mean-field limit of the system (\ref{ODEs} is given by a Vlasov-type PDE coupled with a system of controlled ODEs. The Vlasov-type PDE describes the evolution of the density of human-driven vehicles as a measure and the ODEs describe the controlled behavior of the $M$ AVs. Specifically, the interaction between $M$ AVs and the human-driven vehicles can be modeled using the following system: 
\begin{equation}
			\label{Mean-field_limit}
			\begin{aligned}
			&\dot{y}_i = w_i, \\ 
			&\dot{w}_i = \left(H_1*_1(\mu + \mu_M)+H_2*(\mu+\mu_M)\right)(y_i, w_i)+u_i, \, i\in I_M, \\ 
			&\partial_t \mu + v \partial_x \mu + \partial_{v} \left(\left(H_1*_1 (\mu+\mu_M) + H_2*(\mu+\mu_M) \right)\mu\right)
			=0, 
			\end{aligned}
		\end{equation}
where $(y, w) \colon t \in [0, T] \mapsto (y(t), w(t)) \in \mathbb{R}^M \times \mathbb{R}_{\geq 0}^M$ is the position-velocity vector of the $M$ AVs, $\mu_M$ is defined as in (\ref{atomic_measure_av}) tracking the position and velocity of the $M$ AVs, $H_1 \colon \mathbb{R}\times \mathbb{R}^{+} \to \mathbb{R}$, $H_2\colon \mathbb{R}\times \mathbb{R} \to \mathbb{R}$ are locally Lipschitz convolution kernels with sub-linear growth, $*_1$ is the convolution with respect to the first variable and $\mu \in \mathcal{P}(\mathbb{R} \times \mathbb{R}_{\geq 0})$ is a measure on $\mathbb{R} \times \mathbb{R}_{\geq 0}$ representing the density distribution of the human-driven vehicles in position and velocity. 

The rigorous limit process connecting the finite-dimensional system of ODEs (\ref{ODEs}) to an infinite-dimensional system with coupled Vlasov-type partial differential equation (PDE) and ODEs (\ref{Mean-field_limit}) was proved in~\cite{Fornasier_2014} using Wasserstein distance.

In addition, one can use finite-dimensional hybrid systems to model multi-lane, multi-class traffic dynamics with $M$ AVs and $N$ human-driven vehicles on an open stretch
of the road with $m$ lanes. The hybrid nature of the model is based on the vehicles' continuous dynamics and the discrete events due to the vehicle's lane-changing maneuvers. The mean-field limit of the finite-dimensional hybrid system is an infinite-dimensional hybrid system containing a Vlasov-type PDE with a source term, ODEs, and discrete events caused by the lane-changing behavior of the AVs. For the rigorous derivation of the mean-field limit of the finite-dimensional hybrid system, please see~\cite{GPV21}.


\end{sidebar}

\begin{sidebar}{Optimal control of measure PDEs}
\noindent
by Xiaoqian Gong
\section[Optimal control of measure PDEs]{}\label{sbar-OCMP}
\vspace{-5mm}
\renewcommand{\thesequation}{S\arabic{sequation}}
\renewcommand{\thestable}{S\arabic{stable}}
\renewcommand{\thesfigure}{S\arabic{sfigure}}

\sdbarinitial{I}n this section, our goal is to investigate the \acronym{optimal control problem} (OCP) of multi-class traffic consisting of AVs and human-driven vehicles on a single lane. In applications, it aims to minimize congestion, energy consumption, or travel delays, by adding controls on AVs, rather than controlling all vehicles in the population. 

Let us again assume that we have a small number of $M$ AVs that have a large impact on the vehicle population and a fixed large number of $N$ human-driven vehicles that have a small impact on the vehicle population. Let $T>0$ be a fixed time horizon. We choose controls $u \in L^1((0,T), \mathcal{U})$, where $\mathcal{U}$ is a fixed nonempty compact subset of $\mathbb{R}^{M}$. We model the situation by solving the following finite-dimensional optimization problem:

{\scriptsize
\begin{equation}
\label{eqn_FN}
\begin{aligned}
   & \min \limits_{u \in L^1((0,T), \mathcal{U})}F_N(u)=\int_0^{T} \left\{L(y_N(t), w_N(t), \mu_{N}(t))+ \sum \limits_{i=1}^{M} \frac{|u_{i}(t)|}{M}\right\}\dd t\;,
\end{aligned}
\end{equation}
}
\noindent where $L(\cdot)$ is a suitable continuous map in its arguments, $\mu_N$ is the atomic probability measure tracking the positions $x$ and velocities $v$ of the $N$ human-driven vehicles as defined in (\ref{eqn: mu_N}), and the position-velocity vectors $(y_N,w_N)$ and $(x,v)$ satisfies the dynamics (\ref{ODEs}) with given initial datum $(x(0), v(0), y_N(0), w_N(0)) = (x_0, v_0, y_{N,0}, w_{N_0}) \in \mathbb{R}^N \times \mathbb{R}_{\geq 0}^N \times \mathbb{R}^M \times \mathbb{R}_{\geq 0}^M$ and control  $u \in L^1((0,T), \mathcal{U})$. Note that we added the subscript $N$ to the AVs' position-velocity vector $(y,w)$ indicating the dependence of the AVs' positions and velocities on the number of human-driven vehicles $N$. The existence of optimal control for the finite-dimensional optimization problem (\ref{eqn_FN}) was proved in~\cite{FPR2014}. 

The mean-field limit of the finite-dimensional system (\ref{ODEs}) was given by system (\ref{Mean-field_limit}) coupled with Vlasov-type PDE and ODEs when the number of human-driven vehicle goes to infinity, that is, $N \to \infty$. Correspondingly, we introduce the following infinite-dimensional optimization problem:

{\scriptsize
\begin{equation}
\label{eqn_F}
\begin{aligned}
   & \min \limits_{u \in L^1((0,T), \mathcal{U})}F(u)=\int_0^{T} \left\{L(y(t), w(t), \mu(t))+ \sum \limits_{i=1}^{M} \frac{|u_{i}(t)|}{M}\right\}\dd t\;,
\end{aligned}
\end{equation}
}
\noindent where $L(\cdot)$ is a suitable continuous map in its arguments, $(y,w, \mu)$ is the unique solution to system (\ref{Mean-field_limit}) with given initial condition $(y^0, w^0, \mu^0) \in \mathbb{R}^{M} \times \mathbb{R}_{\geq 0}^M \times \mathcal{P}(\mathbb{R} \times \mathbb{R}_{\geq 0})$ ($\mu^0$ is compactly supported) and control $u \in L^1((0,T), \mathcal{U})$. 

It turns out that the cost functional $F_N$ in (\ref{eqn_FN}) $\Gamma-$ converges to the functional $F$ in (\ref{eqn_F}) as $N \to \infty$. This leads to the existence of optimal controls for the infinite-dimensional OCP (\ref{eqn_F}). In fact, the solutions of the finite-dimensional OCP (\ref{eqn_FN}) converge to the optimal controls for the infinite-dimensional OCP (\ref{eqn_F}). For more details, please see~\cite{FPR2014}. 

Furthermore, for multi-lane and multi-class traffic dynamics, we can study the mean-field limit of an OCP of a finite dimensional hybrid systems, which is given by an OCP of an infinite-dimensional hybrid system. The existence of optimal control for the OCP associated with the infinite-dimensional hybrid system is again due to a $\Gamma-$ convergence result. For more detailed discussion, please see~\cite{gong2021meanfield}.

 
\end{sidebar}






\begin{sidebar}{MPC controller}
\noindent
by Fangyu Wu
\section[MPC controller]{}\label{sbar-MPC}
\vspace{-5mm}
\renewcommand{\thesequation}{S\arabic{sequation}}
\renewcommand{\thestable}{S\arabic{stable}}
\renewcommand{\thesfigure}{S\arabic{sfigure}}

\sdbarinitial{S}top-and-go waves, characterized by periods of motion followed by abrupt halts, present significant challenges in traffic control and vehicular automation.
To address this issue, we consider Model Predictive Control (MPC), a popular optimal control method~\cite{borrelli2017predictive}.
The appeal of MPC lies in its ability to translate a typically discrete-time control task into a finite-dimensional optimization problem.

At each discrete time step $i$, the optimization problem obtains an estimate of the initial state of the plant.
It then computes an optimal sequence of control over a specified planning horizon.
Only the first control sequence is dispatched for actuation, with this cycle repeating until task completion.

We denote the state and input of the plant as $\bm{x} \in \mathbb{R}^{n}$ and $\bm{u} \in \mathbb{R}^{m}$ respectively. 
The discrete-time dynamics is expressed as $\bm{x}_{i+1} = f(\bm{x}_{i}, \bm{u}_{i})$, where $f : \mathbb{R}^{n} \times \mathbb{R}^{m} \rightarrow \mathbb{R}^{n}$.
The initial state, $\bm{x}_{0}$, equals $\bm{x}_{\text{init}}$.
The control task imposes the following state and actuation constraints: $\bm{x} \in \mathcal{X}, \bm{u} \in \mathcal{U}$.

The objective of MPC is to minimize the cost function $l = \sum_{i=0}^{N-1}\ell(\bm{x}_{i}, \bm{u}_{i}) + \ell_{f}(\bm{x}_{N})$.
This function measures the cumulative effect of state and control inputs, with $\ell, \ell_{f} : \mathbb{R}^{n} \times \mathbb{R}^{m} \rightarrow \mathbb{R}$.
Analytically, the MPC entails to solving the following optimization at each time step:
\begin{mini}|s|
    {\bm{x}_{i}, \bm{u}_{i}}{\sum_{i=0}^{N-1}\ell(\bm{x}_{i}, \bm{u}_{i}) + \ell_{f}(\bm{x}_{N})}
    {}{}
    \addConstraint{\bm{x}_{0} = \bm{x}_{\text{init}}}{}
    \addConstraint{\bm{x}_{i+1} = f(\bm{x}_{i}, \bm{u}_{i}),}{\quad i = 0, \dots, N-1}
    \addConstraint{\bm{x}_{i} \in \mathcal{X},}{\quad i = 1, \dots, N}
    \addConstraint{\bm{u}_{i} \in \mathcal{U},}{\quad i = 0, \dots, N-1.}
\end{mini}
Upon each iteration, the MPC accepts $x_{\text{init}}$ as input and generates $u_0$ as output.
This feedback control process continues in a receding horizon manner until the task is terminated.

For the CIRCLES project, our team has developed an MPC that adopts a linearly constrained quadratic programming formulation, as demonstrated in~\cite{wu2022hierarchical}.
This control method primarily focuses on the longitudinal dynamics of the ego vehicle, defined as the vehicle controlled by the algorithm.
The state comprises the vehicle's position and velocity, and the control input is its acceleration.

The imposed state constraints ensure that the ego vehicle neither collides with nor overtakes the predicted position of the lead vehicle.
They also enforce a maximum road speed limit. Concurrently, input constraints set upper and lower bounds on acceleration. 
The MPC's objective is to minimize the sum of the $L^2$ norm on acceleration.
This objective leads to a standard convex quadratic programming problem, which is solvable by widely available solvers.

The primary challenge associated with employing MPC for wave attenuation lies in accurately predicting the lead vehicle's longitudinal position across a substantial planning horizon. To optimally smooth traffic waves, it is imperative that this planning horizon aligns with both the spatial and temporal scales of the wave.

To overcome this challenge, our approach incorporates a realtime map service for long-term prediction and leverages instantaneous acceleration extrapolation for short-term forecasting. This dual predictive strategy serves two critical functions: the long-term predictions guide the ego vehicle to effectively dampen the traffic waves, while the short-term forecasts act as safeguards, preventing potential collisions with the lead vehicle.


\end{sidebar}

\begin{sidebar}{Optimized Vehicle Trajectory}
\noindent
by Arwa AlAnqary
\section[Optimized Vehicle Trajectory]{}\label{sbar-OC}
\vspace{-5mm}
\setcounter{sequation}{0}
\renewcommand{\thesequation}{S\arabic{sequation}}
\renewcommand{\thestable}{S\arabic{stable}}
\renewcommand{\thesfigure}{S\arabic{sfigure}}

\sdbarinitial{T}o gain more insights into the car-following task and to create a baseline for performance benchmark of control algorithms, we propose an optimal control formulation of the problem. We consider a mixed-autonomy platoon of vehicles driving on a single lane following a leader with pre-specified trajectory over a fixed interval $[0, T]$. The aim is to find the optimal control signal for all automated vehicles in the platoon to minimize an the platoon's energy consumption. 
 
\subsection{Optimal Control Problem}
We consider a mixed autonomy platoon of $M$ AVs, $N$ \emph{human vehicles} (HVs), and a leader vehicle. 
The AVs are controlled in their acceleration, the HVs’ acceleration is governed by a \emph{car-following model} (CFM) \(A \colon \mathbb{R}^{2}  \times \mathbb{R}_{\geq 0}^2 \to \mathbb{R}\), and the leader trajectory of the leader is specified by its position $x_\ell(t)$ and velocity $v_\ell(t)$. 
We index the vehicles in the platoon from front to back with $i = 0$ being the leader vehicle.
Let $\mathbf{u}(t), \mathbf{x}(t), \mathbf{v}(t), \in \mathbb{R}^{N+M}$ be the control, position, and velocity vectors, respectively. 
Given initial value vectors $\mathbf{x}_{0}$ and $\mathbf{v}_{0}$, the platoon dynamics are governed by the following system of ODEs
\begin{equation}
\small
\begin{aligned}
\dot{\mathbf{x}}_{i}(t) &=\mathbf{v}_{i}(t),& i \in I \setminus \{0\}\;,\\
\dot{\mathbf{v}}_{i}(t)&= A \big(\mathbf{x}_{i}(t) ,\mathbf{x}_{i-1}(t),\mathbf{v}_{i}(t),\mathbf{v}_{i-1}(t)\big) &i\in I_h \;,\\
\dot{\mathbf{v}}_{i}(t)&= \mathbf{u}_i(t) &i\in I_a\;,\\
\end{aligned}
\label{eq:dynamics}
\end{equation}
where $I = \{0, 1, \dots, M+N\}$, $I_{a}$ and $I_{h}$ are the sets of indices of the AVs and HVs, respectively. 

We use Bando-follow-the-leader CFM~\cite{bando1995dynamical} for the HVs in the platoon. The model describes the acceleration of the vehicle as a function of its space gap, velocity, and relative velocity. For parameters $\alpha$, $\beta$, $k$, $d$, and car length $l$, we have 
\begin{align}
\label{eq:bando-ftl}
    A(x_\ell, x, v_{\ell},v)= \alpha\big(V(x_\ell-x-l)-v\big)+\beta\tfrac{v_\ell-v}{(x\ell-x-l)^{2}}\;, 
\end{align}
where 
\begin{align}
    V(h) = v_{\max} \tfrac{\text{tanh}(k h - d) + \text{tanh}(l + d)}{1 + \text{tanh}(l + d)}\;. 
\end{align}

Next, we define the set of admissible controllers as the set of functions $\mathbf{u}: [0, T] \mapsto \mathbf{R}$ that satisfy certain conditions: (1) the controllers can only be applied to the AVs, and they are bounded; (2)  the AVs can not drive backwards; (3) the AVs remain within an allowable space gap envelope. 

Lastly, we define the the objective functional of the \emph{optimal control problem} (OCP) as the $L^2$ norm of the acceleration of all the vehicles in the platoon. We use this as a simple proxy of the fuel consumption of the vehicles. 

Based on the above, we formulate the following OCP 
\begin{equation}\label{eq:opt_problem_energy}
    \inf_{\boldsymbol{u}} \int_{0}^{T} \sum_{i \in I_a} \boldsymbol{u}_i^2(t) + \sum_{i \in I_h}\big(\text{A}(\boldsymbol{x}_{i-1}(t), \boldsymbol{x}_i(t), \boldsymbol{v}_{i-1}(t), \boldsymbol{v}_{i}(t))\big)^{2} \dd t\;,
\end{equation}
where $(\boldsymbol{x}, \boldsymbol{v})$ satisfies Equation (\ref{eq:dynamics}) and $\forall i \in I_a,\ \forall t\in[0,T]$
\begin{equation}
\begin{aligned}
h_{\min}\boldsymbol{v}_i(t) + d_{\min}&\leq \boldsymbol{x}_{i-1}(t) - \boldsymbol{x}_i(t) - l\;,\\
h_{\max}\boldsymbol{v}_i(t) + d_{\max}&\geq \boldsymbol{x}_{i-1}(t) - \boldsymbol{x}_i(t) - l\;,\\
 \boldsymbol{v}_i(t) &\geq 0\;,
\end{aligned}
\end{equation}
where $h_{\min}$ and $h_{\max}$ are bounds on the allowable time gap, and $d_{\min}$ and $d_{\max}$ are the allowable minimum and maximum space gap at zero velocity. 

To solve this optimization problem we parameterize the controls using piece-wise constant functions. This renders a finite dimensional optimization problem that we solve by means of gradient descent. 
%
We compute the analytical gradients of the problem using the adjoint formulation. For further details see \cite{cdcoptimal}.

\subsection{Numerical Simulation} We apply the proposed approach to a platoon with one AV following an empirical leader trajectory. This trajectory exhibits stop-and-go waves. We show the simulation results in Figure \ref{fig:oc:simple}. By introducing one AV, we achieve up to $10\%$ reduction in the fuel consumption compared to the baseline of full human-driven platoon. We apply the same approach sequentially to multiple AVs in a larger platoon. In Figure \ref{fig:oc:tsd} we show the time-space diagrams for the baseline platoon compared to one with 8 AVs which achieves $24\%$ reduction in the fuel consumption. 

\sdbarfig{\label{fig:oc:simple}\includegraphics[width=19.0pc]{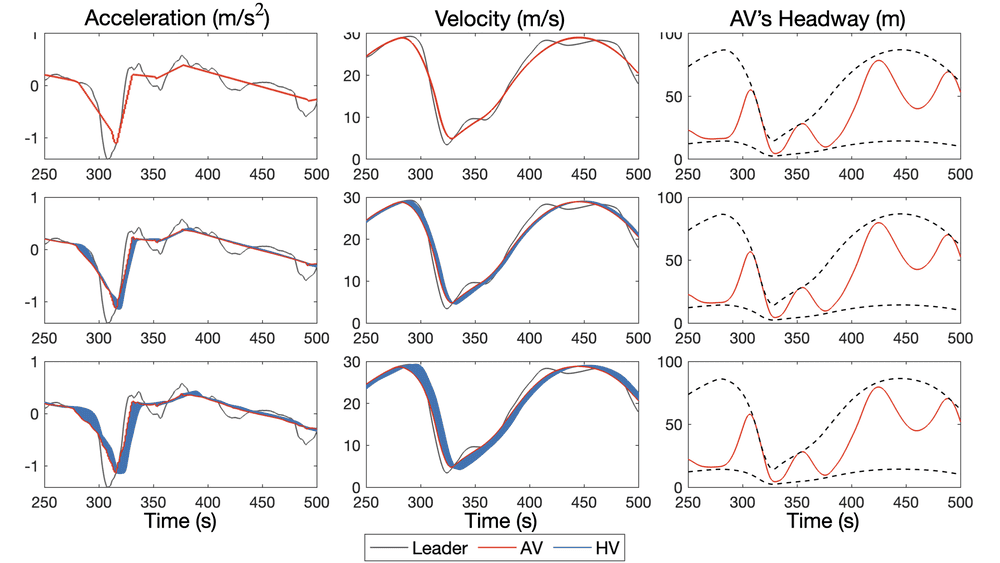}}{Trajectories of different platoons following the considered leader. Platoon size: 0 HVs (top row), 10 HVs (middle row), and 20 HVs (bottom row). The dashed lines in the space gap plots represent the feasible space gap profile. By introducing one AV behind the leader, we can achieve up to 10\% reduction in the energy consumption compared to the fully human-driven platoon.}
\end{sidebar}

\renewcommand{\thesequation}{S\arabic{sequation}}
\renewcommand{\thestable}{S\arabic{stable}}
\renewcommand{\thesfigure}{S\arabic{sfigure}}

\begin{sidebar}{\continuesidebar}
\sdbarfig{\label{fig:oc:tsd}
    \includegraphics[width=19.0pc]{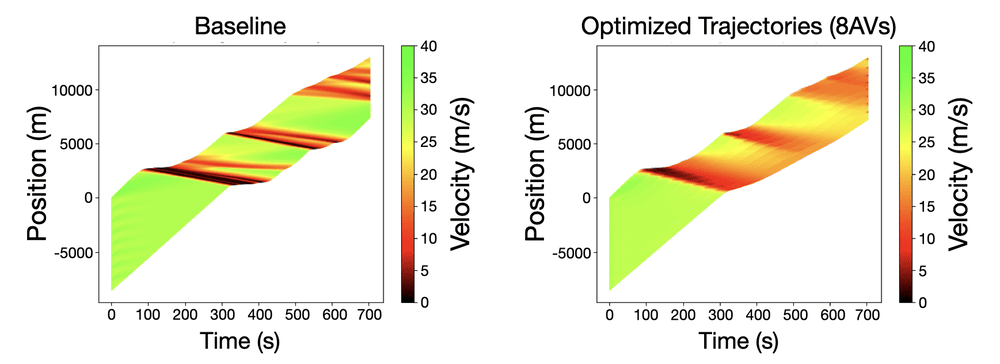}}{Time-space diagrams showing the effect of introducing 8 AVs with trajectories optimized using the proposed approach (right) and comparing it with a baseline of a full human-driven platoon (left). The introduction of the AVs dampens the propagation of stop-and-go waves that appear in the leader's trajectory.  It also achieves $24\%$ reduction in the fuel consumption compared to the baseline. }

\end{sidebar}

\begin{figure}
    \centering
    \includegraphics[width=\columnwidth]{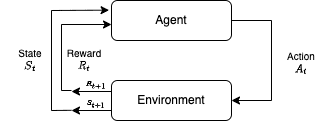}
    \caption {The Markov Decision Process that RL is based on. An agent exists in an environment and repeatedly chooses action based upon a state and receives rewards, which then informs the agent on the value of the state and action pair.}
    \label{fig:mdp}
\end{figure}

\subsection{Speed Planner}\label{sec:SP}

Based on the hierarchical framework, {Figure \ref{fig:planner} indicates the implementation of the speed planner we tested in the MVT.  As introduced in Table~\ref{table:data_sources}, INRIX and probe vehicle data are posted to the central database in different frequencies: each vehicle makes its post approximately every 1 second, and a server-side process inserts new data from INRIX approximately every 60 seconds. It is important to note that while the INRIX data provides a single speed across all lanes, the AV pings provide lane-level speed information.

The sequence of events of a Speed Plan publication can be summarized as follows:
\begin{enumerate}
    \item Each new INRIX update is combined with historical INRIX data, and fed into the prediction module.
    \item All vehicle observations from the previous 60 seconds are fetched, and fused with the INRIX prediction, to obtain a lane-level traffic state estimate. 
    \item The lane-level TSE is smoothed with the forward-kernel average.
    \item Bottleneck identification is performed with the smoothed lane-level TSE.
    \item If a standing bottleneck is identified in the lane, a deceleration region is prescribed as a buffer segment. For all other regions, the smoothed lane-level TSE is used as the lane-level Speed Plan.
    \item Publish the Speed Plan for all lanes.
\end{enumerate}

The subsequent sections delve deeper into the intricacies of the TSE Enhancement and Target Design, providing a comprehensive understanding of the steps outlined above. Wang et al. \cite{WangSpeedplanner} provides in-depth methodology description of the Speed Planner.

\begin{figure}[h!b]
\centering
\includegraphics[width=0.50\textwidth]
{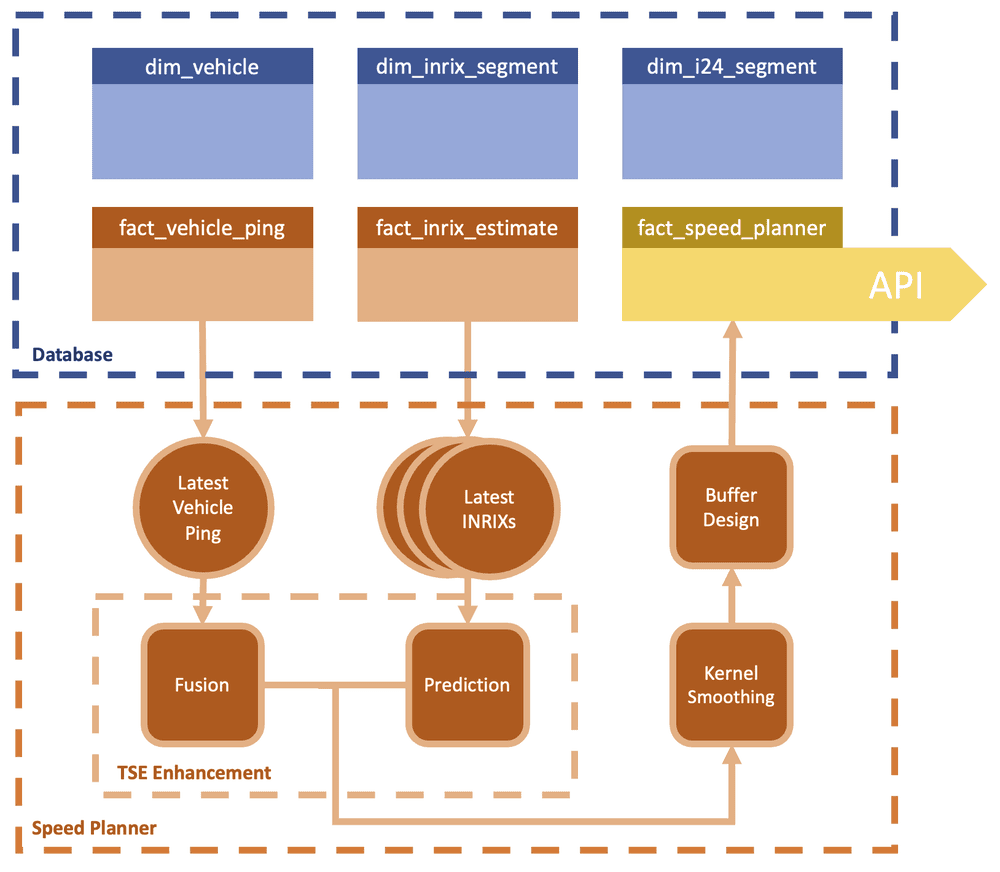}
\caption{\textbf{Data Pipeline and Major Function Modules of Speed Planner} At the beginning of each update, the Speed Planner extracts a combination of macroscopic TSE and vehicle observations from the corresponding factual tables (\texttt{fact\_inrix\_estimate}, \texttt{fact\_vehicle\_ping}) in the database to calculate the target speed profile. The raw TSE is used as the input of the prediction module, of which the output is fused with vehicle observations. The fusion is then smoothed and used in the buffer design module, of which the output is saved into the database (\texttt{fact\_speed\_planner}) and published as the target speed profile.}
\label{fig:planner}
\end{figure}

\subsubsection{TSE Enhancement}
The TSE enhancement module comprises two main components: the INRIX prediction module and the data fusion module. The prediction module's primary role is to minimize the latency issues associated with INRIX real-time data, especially when applied to vehicular control, as opposed to its standard use for general traffic insights. The fusion module complements this by integrating real-time data from our system's probe vehicles, enabling a more detailed, lane-level TSE with enhanced time-space precision. A detailed mathematical explanation of the TSE enhancement module for MVT implementation is available in~\cite{WangSpeedplanner}.

To achieve a lane-level TSE with superior spatial detail, we further segment the INRIX data into smaller units. After generating the INRIX prediction, it's merged with real-time data from the system's controlled vehicles. Given that each AV communicates with the server at a rate of 1Hz, we receive 60 ping records from each vehicle for every Speed Plan generation. These records help determine the average speed of each vehicle over the previous update period, which then updates the TSE for the respective sub-segment. Assuming that our drivers consistently stick to their designated lanes (a behavior largely confirmed through systematic data review), we can generate lane-specific TSE estimates by combining vehicle data with the broader INRIX speed data. When it comes to data fusion, we give precedence to data from our vehicles over INRIX data, considering the inherent characteristics of both sources. Our vehicles gather and relay perception data through a system that's both observable and controllable, with quantifiable error and latency as detailed in~\cite{richardson2023analysis}. In contrast, the INRIX real-time API employs averaging techniques to forecast over a given period. This method might introduce inaccuracies at specific points within that timeframe, which could impact our control execution. The INRIX system, in essence, operates as a somewhat opaque system, with its error and latency aspects largely inferred from provider descriptions. In this paper, we adopt the following notation to represent the discrete TSE:
\begin{equation}\label{E:inrix_seg}
\{(x_j,\Bar{v_j}), j \in \mathcal J\},
\end{equation}
where $j$ represents the index of road segments, $x_j$ is the postmile coordination of the central of road segment $j$ and $\Bar{v_j}$ is the average speed of the corresponding road segment $j$. Wang et al. \cite{WangSpeedplanner} details the procedure of data fusion applied in MVT.


\subsubsection{Target Design} 
This section introduces the Target Design's main modules, namely the kernel-based smoothing and the learning-based buffer design. The kernel smoothing processes the enhanced TSE at each time step using a chosen kernel to improve the fuel consumption caused by the shockwave in a high-density traffic flow. The buffer design utilizes Reinforcement Learning (RL) to form a buffer area upstream of the standing bottleneck with the goal of improving throughput at the bottleneck. The target speed suggested by the RL is employed in a mathematical model of traffic, represented by a strongly coupled Partial and Ordinary Differential Equation (PDE-ODE). The outcome of this mathematical model is an identification of traffic density $(t, x) \mapsto \rho(t, x)$. The kernel smoothing then receives this information for learning the velocity of the next time step. 

In kernels smooth module, we rely on enhanced TSE data to synchronize the driving speeds of automated vehicles. In particular, vehicles are assigned target speed profiles contingent on traffic state information, which is shared and common among all AVs. 

At any fixed time step $t$, the desired speed profile $\Bv:\RR_+ \times \RR \to \RR_+$ is extracted from enhanced TSE utilizing kernel methods. First, we preprocess the sparse TSE data by interpolating the discrete data pairs $(x_i, \Bar{v_i})$ to a continuous speed profile $(t, x)  \in \RR_+ \times \RR \mapsto v(t, x)$, as an approximation of the average speed of a higher granularity traffic at a position $x$ and at time $t$. Then, for any fixed time $t = t_\circ$ we obtain the desired speed by applying a kernel function $K(\cdot)$ at a position $x = x_\alpha$: 
\begin{equation}
    \Bv(t_\circ, x_\alpha) = \frac{\int_{x=x_\alpha}^{x_\alpha+w} K(x_\alpha, x) v(t_\circ,x)dx}{\int_{x=x_\alpha}^{x_\alpha+w} K(x_\alpha, x)dx}, \label{eq:desire_kernel}
\end{equation}
where $w$ is the width of the estimation window. Many different kernel functions, such as Gaussian kernel, Triangular kernel, Quartic kernel, Uniform kernel etc., can be chosen. For the purposes of this paper, we consider a uniform kernel, the simplest of such mapping. The desired speed profile at a position $x_\alpha$ is accordingly defined as:
\begin{equation}
    \Bv(t_\circ, x_\alpha) = \frac{\int_{x=x_\alpha}^{x_\alpha+w} v(t_\circ, x)dx}{w}. \label{eq:desire}
\end{equation}
 
When  human drivers observe a gap between their vehicle and the one preceding, they tend to accelerate to close the distance. Our proposed desired speed profile aims to slow down \emph{in advance}, although not excessively, to create a gap from the preceding vehicle. This approach takes into account the information provided by the TSE, which indicates the presence of congestion in the nearby downstream area. The proposed desired speed profile is adaptive to traffic states and offers relative robustness, as it only requires one parameter, $w$, to tune.

\begin{pullquote}
    Congestion induces a reduction in system efficiency due to the uneven distribution of traffic density in the time-space domain.
\end{pullquote}

For buffer design, we consider the interval $\mathcal I\subset \RR$ as the region of interest. In addition, we consider a subregion $\mathcal I_c \subset \mathcal I$ as a congested area. The idea is to determine the controlled vehicle target speed at a time step $t_\circ$, denoted by $\Bu(t_\circ, x)$, such that the density $\rho(t, x)$ for $x \in \mathcal I_c$ and $t \ge t_\circ$ is distributed uniformly through the region $\mathcal I$. Determining the controlled vehicle's target speed will be done in the following steps: 
(i) Designing a target speed $\Bu(t_\circ, x)$, given the input $\Bv(t_\circ, x)$ from kernel smoothing step, 
(ii) identifying the density $(t, x)\in \RR_+ \times \mathcal I \mapsto \rho(t,x)$, given the target speed of the controlled vehicle, employing a strongly coupled PDE-ODE model of traffic flow, 
(iii) Evaluation step in which using the density, the speed profile will be updated by smoothing kernel. 

Paper \cite{WangSpeedplanner} provides details of the procedure introduced above.

\subsection{Vehicle Controller}
Recall in the architectural design of the controller, the Vehicle Controller layer features two distinct methods for actuating the vehicle: acceleration-based control and ACC-based control. Here, we will discuss component implementations prepared for both options.

\subsubsection{Acceleration-Based Controller: Base Control}\label{sec:microaccel}
The acceleration-based controller requires a baseline controller to use under normal and emergency operating conditions, and a lane-change recovery controller to behave smoothly when cut-ins and cut-outs take place.

The base controller is an explicit, mathematically-defined controller designed to reach and keep an ideal target speed without being trapped in the stop-and-go wave. These two goals are somewhat antagonistic: a naive approach consisting in following the ideal target speed (unless it is unsafe to do so) may result in braking at the same time and with a comparable amplitude as the leading car when caught in the wave. 

To address this, the base controller has an anticipation mechanism based on a paradigm that can be summarized as ``act swiftly but slightly,'' which reduces variability in acceleration. It is comprised of three key components: \emph{target}, \emph{anticipation}, and \emph{safety}. The target component infers an ideal target speed for each specific AV based on inputs from the Speed Planner and from the local data available. The anticipation component is an MPC module that aims to anticipate the leader's behavior based on its current acceleration. This enhances the controller's ability to follow the target speed effectively. The safety component is a safety module, that has priority over all the other components and ensures that the AV remains safe at all times. 

The mathematical expression of the commanded acceleration is given by
\begin{equation}
    a_{\text{cmd}} = \min(a_{\text{safe}},a_{\text{target}},a_{\text{MPC}})\;,
\end{equation}
where $a_{\text{safe}}$, $a_{\text{target}}$, and $a_{\text{MPC}}$ correspond to the \emph{safety}, \emph{target}, and the MPC-based \emph{anticipation} components. The safety component is given by
\begin{equation}
\label{eq:asafe}
\begin{split}
a_{\text{safe}}(t) &= -k(v(t)-v_{\text{safe}}(t))+\dot v_{\text{safe}}(t)\;,\\
v_{\text{safe}}(t)&= \sqrt{ 2|a_{\min}|\left(h(t)-s_{0}+\frac{1}{2} \frac{v_{lead}^{2}(t)}{|a_{l,\min}|}\right)}\;,
\end{split}
\end{equation}
where $k$ is a positive parameter, $v$ is the velocity of the ego vehicle, $s_{0}$ is  safety distance, $h$ is the space gap, $v_{lead}$ is the leader vehicle velocity and $a_{l,\min}$ is the minimal possible leader vehicle acceleration (so maximal deceleration), and $a_{\min}$ the minimal possible ego acceleration. 

The target component is defined by
\begin{equation}
a_{\text{target}}(t) = -k(v(t)-v_{\text{target}}(t))\;,
\end{equation}
where $v_{\text{target}}$ is an inferred ideal speed obtained either from the speed planner when available or by integrating local measurements.

The MPC-based anticipation component $a_{\text{MPC}}$ is given by
\begin{equation}
\small
 a_{\text{MPC}}(t)= \left\{
\begin{aligned}
 &a_{\text{min. brake}}(h(t),v(t),v_{lead}(t),a_{lead}(t)),\\
 &\hspace{3.5 cm}\text{ if } P_1>0,\;a_{lead}(t)<0 \\
&a_{lead}v(t)/v_{lead}(t), \\
 &\hspace{2.0 cm}\text{ if } P_1\leq 0 \text{ and } P_2\geq 0,\; a_{lead}(t)<0\\
&a_{lead}-\frac{(v(t)-v_{lead}(t))^{2}}{2(h(t)-s_{0})},\\
 &\hspace{2.0 cm} \text{ if } P_1\leq 0 \text{ and } P_2< 0,\;a_{lead}(t)<0\\
&a_{lead}-\frac{(v(t)-v_{lead}(t))^{2}}{2(h(t)-s_{0})},\\
 &\hspace{3.5 cm} \text{ if }  P_2< 0,\;a_{lead}(t)\geq 0, \\
&\min(a_{\max}, a_{lead}(t)(1+k_{2}(v_{\text{lead}}(t)-v(t))), 
\\
 &\hspace{3.5 cm}
\text{ otherwise}.
\end{aligned}
\right.
\end{equation}
where, as in \eqref{eq:asafe}, $h$ is the space gap and $s_{0}$ the safety distance, and $k_{2}$ is a positive constant, $a_{lead}$ is the leader vehicle acceleration, $a_{\max}$ is the maximal possible acceleration of the ego vehicle. The terms $a_{\text{min. brake}}$, $P_{1}$, and $P_{2}$ are given by
\begin{equation}
\label{eq:a12}
\begin{split}
&a_{\text{min. brake}}(h(t),v(t),v_{lead}(t),a_{lead}(t)) \\
&=-\left(h(t)-s_{0}+\frac{1}{2} \frac{v_{lead}^{2}(t)}{-a_{lead}(t)}\right)^{-1} \frac{(v(t))^{2}}{2}\;,\\
P_1&= a_{\text{min. brake}}(h(t),v(t),v_{lead}(t),a_{lead}(t))\\
&-a_{lead}(t)v(t)/v_{lead}(t)\;, \\
P_2&= v_{lead}(t)- v(t)\;.
\end{split}
\end{equation}

Further details of this controller can be found in~\cite{HayatMicrocontroller}.

\subsubsection{Acceleration-Based Controller: Lane-Change}
The design of Lane-Change Recovery Controller relies on global traffic information as well as local traffic states, such as the space gap to the leader vehicle and the relative velocity. The event of lane changing of the leader vehicle creates a discontinuity in the observed local state. Such discontinuities can translate to sudden, large jumps in the controller output, causing large jerk values. This sudden jump in the controller output might be necessary to avoid collision (for example, if a vehicle cuts-in in front of the AV and has low relative velocity). However, in many cases, these jumps are byproducts of the discontinuity in the observation and can be avoided without creating additional safety threats. To remedy this effect of lane-changes in such conditions, we designed a simple lane-change handling mechanism that treats the main acceleration-based controller as a black box and makes minimal assumptions about it. Essentially, the mechanism works by detecting the event of lane changing, assessing the safety conditions created by the event, and (when appropriate) smoothing the output of the controller. 

We assume discrete observations with fixed frequency. At a step $k$, we denote the ego velocity $v_k$, the relative velocity $\Delta v_k$, the space gap $s_k$, the output of the main controller $a_k$, and the actual acceleration input to the vehicle $u_k$. If the lane-change mechanism is not active, we have $u_k = a_k$.

In order to design this mechanism, we make a few natural assumptions about the behavior of the underlying controller. We assume that the controller is a continuous and non-decreasing function of the leader's states (that is, the space gap $s_k$ and relative velocity $\Delta v_k$). We also assume that the main controller encodes its own safety measures as the lane-change handling mechanism does not provide any additional safety measures. Beyond these assumptions, the exact form of the main controller is treated as a black box. 

We impose multiple criteria for the lane-change handling mechanism to take effect:
\begin{enumerate}
    \item a lane-change event is detected---this is done by detecting a discontinuity in the space gap to the leader vehicle;
    \item the lane-change event is significant---this is measured by the amount of jerk it causes;
    \item the lane-change event does not cause safety threats---this is measured by the \emph{time to collision} (TTC) (if $\Delta v_k <0$) or time gap (if $\Delta v_k \geq 0$) at the lane-change event.
\end{enumerate} 
When the lane-change mechanism is activated, it smooths the acceleration by taking a convex combination of the main controller output at the current step and the actual acceleration command in the previous step:
\begin{equation}
    u_k = a_k u_{k-1} + (1-a_k) a_k\;. 
\end{equation}
Here $0\leq a_k < 1$ is a smoothing factor where larger values mean smoother acceleration change. Here we consider a time-varying $a_k$ whose value is a non-decreasing function of the relative velocity and time headway at step $k$. The rationale for this modeling choice is to make the smoothing effect stronger the less critical the situation is (that is, larger space gap and relative velocity). The smoothing factor has the following functional form:
\begin{equation}
    a_k = c \cdot \left(f_1\left(\frac{s_k}{ v_k}\right) \right) 
    + (1-c)\cdot f_2(\Delta v_k, s_k)\;. 
\end{equation}
It depends on the time headway through the function 
\begin{align}
    f_1\left(\frac{s_k}{ v_k}\right) = \tanh \left(t^{\star} \cdot \frac{s_k}{ v_k}\right) 
\end{align}
where the parameter $t^{\star}$ is chosen such then the function is close to $1$ when the time headway $\frac{s_k}{\Delta v_k} \geq h_{\text{safe}}$, where $h_{\text{safe}}$ is an acceptable safety time headway. The smoothing factor depends on the relative velocity through the function 
\begin{align}
    f_2(\Delta v_k, s_k) = \begin{cases} 
    \frac{1}{2} \tanh \left(\Delta v^{\star} \cdot \frac{s_k}{|\Delta v_k|}  \right) + \frac{1}{2} & \Delta v_k < 0 \\
    \frac{1}{2} \tanh \left(\Delta v_k  \right) + \frac{1}{2} & \Delta v_k \geq  0 \\
    \end{cases}
\end{align}
where the parameter $\Delta v^{\star}$ is selected such that the function value is close to $0$ when the time to collision $\frac{s_k}{|\Delta v_k|} \leq C_{\text{safe}}$, where $C_{\text{safe}}$ is an acceptable safety time to collision. Concretely, we use $h_{\text{safe}} = 2s$, $C_{\text{safe}} = 4.5s$, $t^{\star} = 1.32$ and $\Delta v^{\star} = 10.3$. 
The parameters $c = 0.75$ is chosen by testing on multiple trajectories and lane-change scenarios. 

Finally, the lane-change mechanism is deactivated when the main controller acceleration becomes close enough to the output acceleration 
\begin{equation}
\left|a_k - u_k\right| \leq \epsilon\;.
\end{equation}
An example of this mechanism is depicted in Figure \ref{fig:lcrc}. In the figure, the dashed red line indicates the lane-change event (cut-in) happening at around 14 seconds. The acceleration profile of the main controller (shown in gray in the bottom left panel) decelerates heavily in reaction to the lane-change.  This event is detected and considered safe by the lane-change recovery controller which remained active for 2.9 seconds during which it smoothed the acceleration profile significantly and removed the unnecessary jerk cause by the lane-change.

\begin{figure}
    \centering
    \includegraphics[width=\columnwidth]{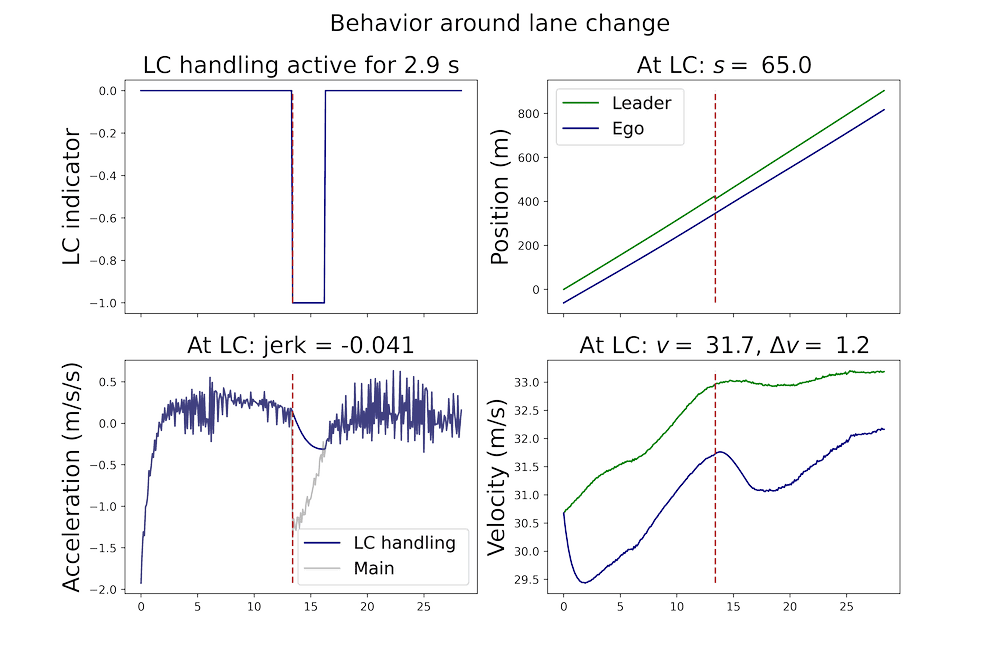}
    \caption{An example of the effect of the lane-change handling mechanism on a real-world trajectory in the event of a cut-in. \textit{Top left}: lane-change event is detected and considered safe so the lane-change recovery controller was active for 2.9s.  \textit{Top right}: a car cuts-in in front of the ego vehicle at a headway of 65m. \textit{Bottom left: } the main controller commanded acceleration drops sharply due to the lane-change event causing a large jerk value, but the lane-change controller smooths out this drop in the acceleration. \text{Bottom right: } The relative velocity is large enough at the lane-change allowing the controller to be active.}
    \label{fig:lcrc}
\end{figure}


\subsubsection{ACC-Based Controller}


The ACC-based controller is the version of the controller that was ultimately deployed on 97 of the 100 vehicles during the final MVT. 
The crucial distinction between the ACC-based controller and the acceleration-based controller introduced in the prior section is the controller's output. The ACC-based controller, rather than providing an acceleration to actuate, provides outputs setpoints for the AV's native ACC system, which controls the AV's longitudinal movements based on those setpoints. We use the stock ACC system's safety assurances and lane-change handling. Thus while this provides a less direct form of control, it is more robust in ensuring the safety and smoothness of the ride. For more information on how the ACC works, please refer to ``\nameref{sbar-ACC}.'' For context on the simulator that was developed for training this algorithm, please refer to ``\nameref{sbar-TS}.''

The ACC-based controller is an RL controller that is trained using Proximal Policy Optimization~\cite{schulman2017proximal}. Elements of training the \acronym{Markov Decision Process} (MDP) problem~\cite{puterman2014markov} are described below:

\noindent\underline{Observation Space}
\begin{itemize}
    \item $v$, velocity of the AV.
    \item $v_s$, target speed given by the Speed Planner.
    \item $l$, a ``minicar'' flag that indicates whether there is a leader vehicle detected, nominally within 80 meters (dubbed as such due to the miniature car icon that appears in the dashboard when the leader is detected).
    \item $s$, the current ACC speed setting 
    \item $g$, the current ACC gap setting
\end{itemize}

\begin{figure}
    \centering
    \includegraphics[width=\columnwidth]{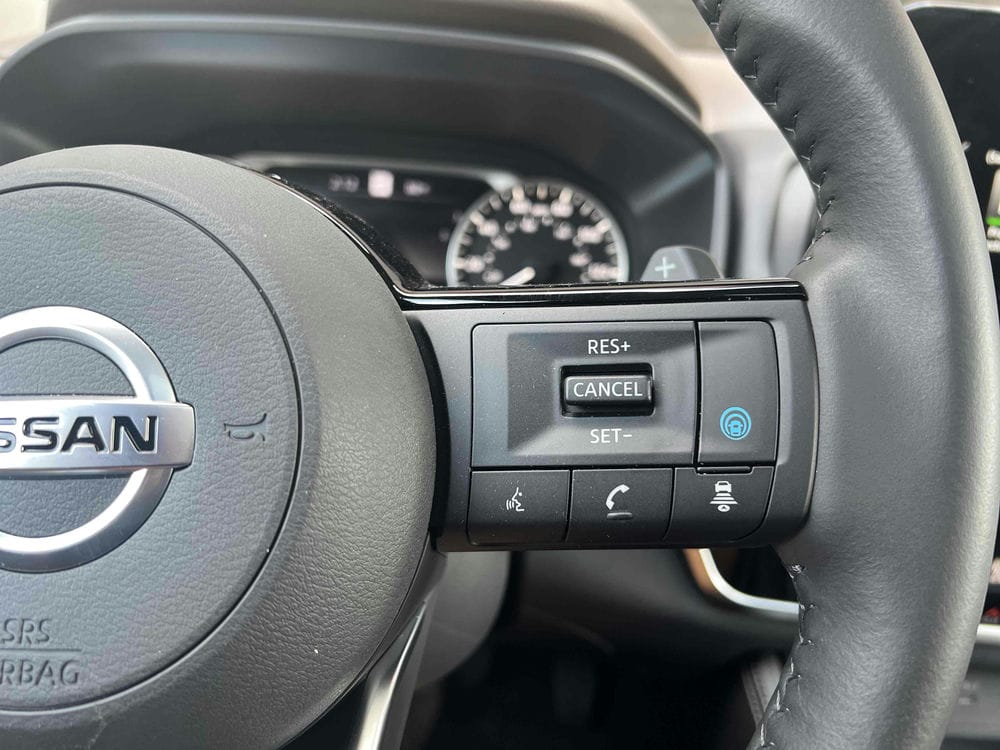}
    \caption{A photograph of a Nissan Rogue's steering wheel buttons that control the vehicle's ACC. The ACC system is turned on by manually pressing the blue icon on the far right. Through our vehicle interfacing efforts, we are able to electronically press the $+$ and $-$ buttons to toggle the ACC speed setting up 1 mph and down 1 mph, respectively, or hold them to increment by 5 mph. The three ACC gap settings are rotated through by pressing the button on the bottom right.}
    \label{fig:acc_buttons}
\end{figure}

\noindent\underline{Action Space}
\begin{itemize}
    \item The requested ACC speed setting, which our onboard computer will realize in a series of button presses (see Figure \ref{fig:acc_buttons}). The speed setting dictates the maximum speed at which the ACC can drive.
    \item The requested ACC gap setting, which the onboard computer will realize with priority over the speed setting. The gap setting takes on three bars between one and three, with each bar indicating a higher allowable gap. Each bar roughly corresponds to constant time gaps of 1.2, 1.5, and 2.0 seconds.
\end{itemize}

In addition, a clipping mechanism was used to ensure controller safety and social acceptability, particularly in the absence of leader state information. The \emph{post facto} lower and upper bounds placed on the speed setting output are based on the average speed of the ego vehicle during the last one second (ten timesteps, with 0 mph and null observations omitted):
\begin{equation}
\begin{split}
v_{\text{lower}} &= \frac{1}{10}\sum_{i=1}^{10} v_i - 15 \text{mph}\;,\\
v_{\text{upper}} &= \frac{1}{10}\sum_{i=1}^{10} v_i + 5 \text{mph}\;,
\end{split}
\end{equation}
where $v_{\text{lower}}$ and $v_{\text{upper}}$ are the lower and upper bounds of the clip, respectively. The final clip is executed as:
\begin{equation}
\begin{split}
v_{\text{clip}} &= \min(\max(v_{\text{action}}, v_{\text{lower}}), v_{\text{upper}})\,\\
v_{\text{final}} &= \min(\max(v_{\text{clip}}, 20 \text{mph}), 73 \text{mph})\;,
\end{split}
\end{equation}
where $v_{\text{final}}$ is the final speed setting command and $v_{\text{action}}$ is the un-normalized speed output, or action, from the neural net. The 73 mph maximum speed accounts for modest speeds in excess of the posted speed limit of 70 mph.

\noindent\underline{Reward Function}
The reward function is:
\begin{equation}
\begin{split}
r_t =& 1 - c_1 a_t^2 - c_2(v_t^{av} - v_t^{sp})^2 +...\\
&-\frac{c_3}{n} \sum_{i=1}^n E_t^i - c_4 \mathbb{1}_{h_t^{av} \leq h_{min} \lor h_t^{av} \geq h_{max}}\;,
\end{split}
\end{equation}
where $c_1-c_4$ are coefficients, $a_t^2$ is an acceleration penalty; $(v_t^{av} - v_t^{sp})^2$ is a squared penalty on the difference between the Speed Planner's suggested speed and the actual speed; $\frac{1}{n}\sum_{i=1}^n E_t^i$ is instantaneous fuel consumption according to the energy models described in ``\nameref{sbar-VEM};'' and the last indicator term is an intervention penalty that is invoked if the space gap is less than the minimum space gap or greater than the maximum space gap.

Further details on the development and design decisions for the state and reward representations and the intermediate and final policies are given in depth in~\cite{JangReinforcement}. 

\begin{sidebar}{Vehicle Energy Models}
\noindent
by Nour Khoudari, Sulaiman Almatrudi, Rabie Ramadan, Joy Carpio, Mengsha Yao, Kenneth Butts, Jonathan W.~Lee, Benjamin Seibold
\section[Vehicle Energy Models]{}\label{sbar-VEM}
\vspace{-5mm}
\renewcommand{\thesequation}{S\arabic{sequation}}
\renewcommand{\thestable}{S\arabic{stable}}
\renewcommand{\thesfigure}{S\arabic{sfigure}}

\sdbarinitial{T}he quantification of the energy demand of the vehicles on the road, given their trajectories, requires vehicle-specific energy models that take as an input the velocity profile $v(t)$ and the road grade profile $\theta(t)$ and output the resulting energy/fuel consumption rate $f(t)$. This project requires the quantification of the energy demand of traffic flow at large, composed of many vehicles; and also the use of reinforcement learning and optimization techniques that minimize (under certain constraints) the energy demand of traffic. Thus, the energy models used should accurately represent different vehicle types on the road and should average out any local non-convexity behavior due to gear switching, to avoid trapping the optimizer in local minima. For that purpose, we use energy models derived from a systematic model-reduction procedure to generate simple fitted models. The procedure starts from the fidelity software Autonomie~\cite{autonomie}, for a number of vehicles, each of which represents a typical average vehicle of a given class.
\section{Vehicle Portfolio} 
To capture the diversity and prevalence of different vehicle types on US roads we select a representative group of vehicle classes on which we apply the model-reduction process to derive their corresponding simplified energy models. Those vehicle classes are divided into two categories: (1) light-duty vehicles: compact size sedan, midsize sedan, midsize SUV, and Pickup, and (2) heavy-duty vehicles: Class4PND (Pickup and Delivery) and Class8Tractor. Each vehicle model represents a class of vehicles that have comparable weight (with load assumed half full) and fuel consumption characteristics~\cite{Argonne}.
\section{Autonomie and Virtual Chassis Dynamometer (VCD)}
We use the simulation software Autonomie Rev 16SP7~\cite{autonomie} with a library of energy models for several types of vehicles, including a detailed plant and controller model for its components in MATLAB and Simulink, where the blocks and files can be customized. To build our models, we use (i)~physics-based vehicle parameters extracted directly from Autonomie, (ii)~tuned parameters extracted in an automated fashion by running the Autonomie model on test cycles, and (iii)~performance maps computed gear-by-gear on a complete velocity--load phase space of driving by running Autonomie's customized vehicle models on a VCD (those maps are vehicle speed to engine speed, vehicle speed and wheel force to engine torque, and engine speed and torque to fuel rate).
\section{Semi-Principled Energy Models}
We build an energy model that is semi-principled in that it has a physics-based part using Autonomie's extracted physics-based vehicle parameters, but it also relies on the maps obtained from the VCD. Gear scheduling in this model is based on choosing the feasible gear that yields the minimal fuel consumption, and the torque converter bypass clutch is assumed open in the first gear only. In contrast to Autonomie that considers hysteresis effects, this model yields the fuel consumption rate (and other outputs) as a direct function of instantaneous velocity $v$, acceleration $a$, and road grade $\theta$.
\section{Simplified Energy Models}
A further model-reduction step is conducted by fitting the semi-principled models into simplified models. Those models have a simple polynomial structure that can easily be integrated into optimization and control problems, yet they are highly accurate. The fuel consumption rate function is
\begin{equation}
    f(v,a,\theta) = \max\left\{\beta,~C(v) + P(v)a + Q(v) a_+^2 + Z(v)\theta\right\}\;,
\end{equation}
where $\beta$ is the minimum fuel rate set to be zero or a positive constant depending on the fuel cut criteria, $a_+=\max(-\frac{P(v)}{2Q(v)},a)$, $C(v) = c_0 + c_1 v + c_2 v^2 + c_3 v^3$, $P(v) =  p_0  + p_1 v + p_2 v^2$, $Q(v) =  q_0 + q_1 v$, and $Z(v) =  z_0 + z_1 v + z_2 v^2$. 

In the above functions, $c_0$ ensures that fuel is being consumed at idle, the $c_1$, $c_2$, and $c_3$ terms can be interpreted as fuel consumed due to friction and air-drag, the $P(v)$ term yields fuel demand due to non-zero accelerations, the $Q(v)$ term captures the super-linear trend of fuel rate with respect to $a$, and the $Z(v)$ term captures the fuel consumed due to road grade, with the $z_1$ term playing the role of the weight force exerted at $\theta$. 
Both types of models are validated, for all different vehicle types, against Autonomie models as the ground truth on standard EPA drive cycles~\cite{drivecycles} for flat roads and constant road grades drive cycles, and the results showed that the models are highly accurate (within 4\% for zero road grades)\cite{KhoudariEnergy}.

\end{sidebar}


\section{Candidate Controller Selection}
In the months leading up to the MVT open road test, our team developed a multitude of candidate controllers, as seen in previous sections and sidebars. In order to pare down the list of candidate controllers to be considered for real-world deployment in November 2022, various (Speed Planner, ACC-based Vehicle Controller)-combinations were assessed over a range of simulation scenarios. The simulation and testing framework follows from \cite{lee2021integrated} and features an updated simulation methodology described in ``\nameref{sbar-TS}.''

Simulation scenarios included shockwaves, bottlenecks, and freeflow. Shockwave and freeflow scenarios follow precisely from the simulator described in ``\nameref{sbar-TS}'' and only differ in the leader trajectory (one being stop-and-go, and the other being all high speed). The bottleneck scenario features a dynamically imposed speed limit in a spatial region of the domain. The speed limit is inversely related to the vehicle density of the region, and the scaling parameter is tuned to historically observed speeds in the I-24 region. The bottleneck region is meant to model a weaving area where a close proximity of on-ramps and off-ramps results in increased lane-change frequency.

The \acronym{key performance indicators} (KPIs) used to assess the controllers include fuel economy, throughput, and network speed. Fuel economy is the overall  miles per gallon of all vehicles, and it is computed by applying the energy model (see ``\nameref{sbar-VEM}'') and computing
\begin{equation}
    \text{fuel economy} = \frac{\sum_{t=0}^T\sum_{i=1}^n E_t^i}{\sum_{i=1}^n x^i}\;,
\end{equation}
where the numerator is the total fuel consumed by all vehicles over all simulation time, and the denominator is the total distance traveled by all vehicles. The throughput is measured by counting the number of vehicles crossing various positions along the highway, normalized by time. For the shockwave and freeflow simulations, this is taken to be the straight average of the time-average of five equi-spaced measurement locations. For the bottleneck, this is taken to be the steady-state (or final value) at the measurement location just downstream of the bottleneck region. The network speed KPI is defined as the total distance traveled by all vehicles divided by the total driving time of all vehicles.

At the time of our self-imposed ``controller freeze'' (that is, the time at which substantial changes to controllers are no longer allowed), there were two Speed Planner variants and 12 Vehicle Controller variants. KPI performance for each of these controllers is shown in Figure~\ref{fig:freeze_evals}. The Speed Planner variants are kernel smooth (just the first three steps described in Speed Planner) and kernel smooth with RL buffer (all steps described in Speed Planner). The Vehicle Controller variants include:
\begin{itemize}
    \item \textbf{Simple}: a hand-designed logic-based controller that largely adheres to the Speed Planner suggestion.
    \item \textbf{MicroAccel}: the main controller described in Acceleration-Based Controller with modifications to output ACC set points. This controller itself has six variants with different parameter choices.
    \item \textbf{RL}: the RL controller described in ACC-Based Controller. This controller has five variants with different training meta-parameter choices.
\end{itemize}

In the shockwave scenario, an ideal controller should maximize fuel economy improvements and remain neutral on throughput and network speed. In the bottleneck scenario, an ideal controller should maximize throughput improvements and optionally improve or stay neutral on the other two KPIs. In the freeflow scenario, an ideal controller will not worsen on any of the KPIs. Given these criteria, we determined that the HybridRL Vehicle Controller paired with the Kernel Smooth with RL Buffer Speed Planner presented the ideal combination for the ACC-Based vehicles.

\begin{figure}
  \centering
  \includegraphics[width=0.8\columnwidth]{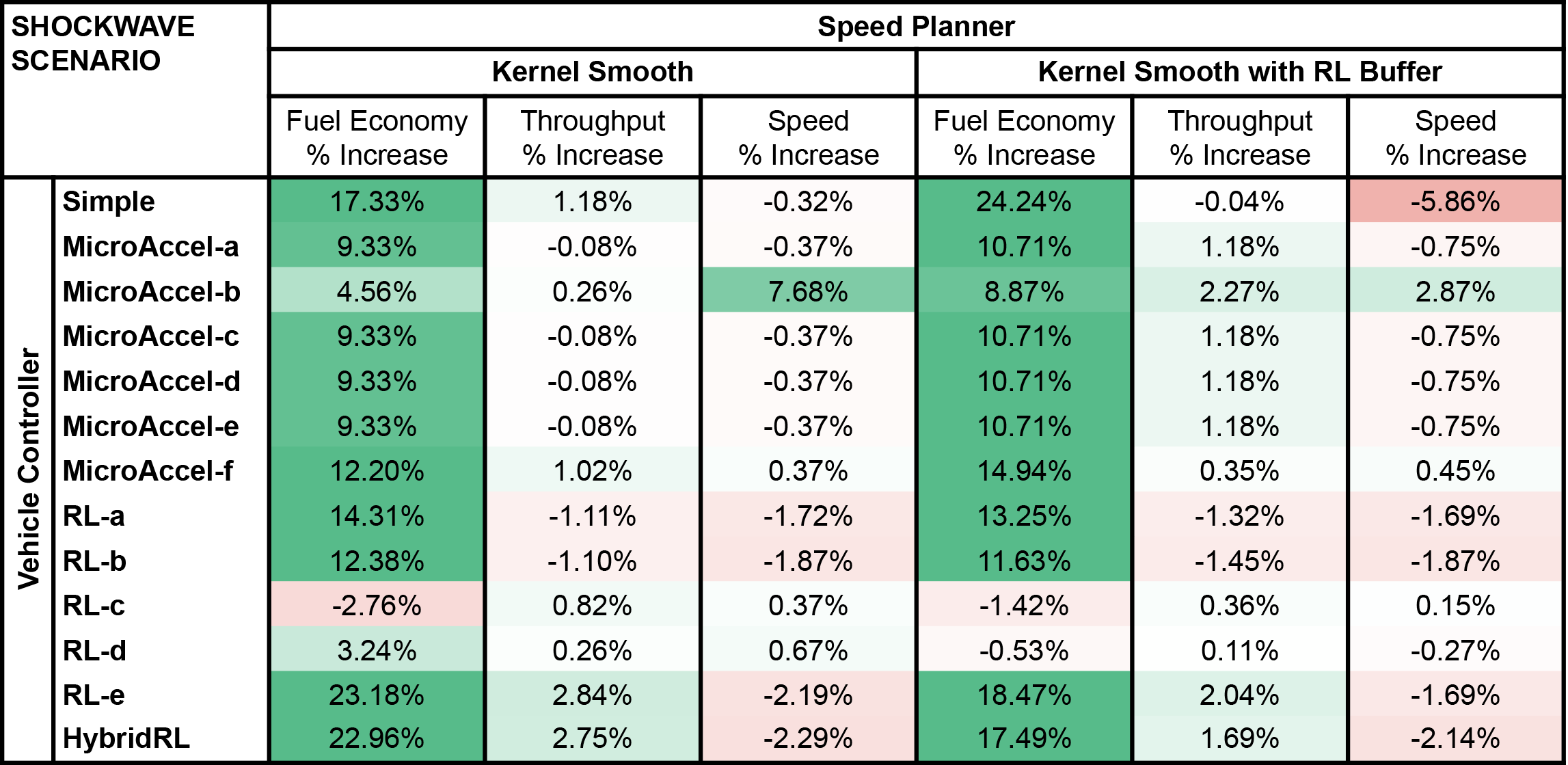}
  \\[\smallskipamount]
  \includegraphics[width=0.8\columnwidth]{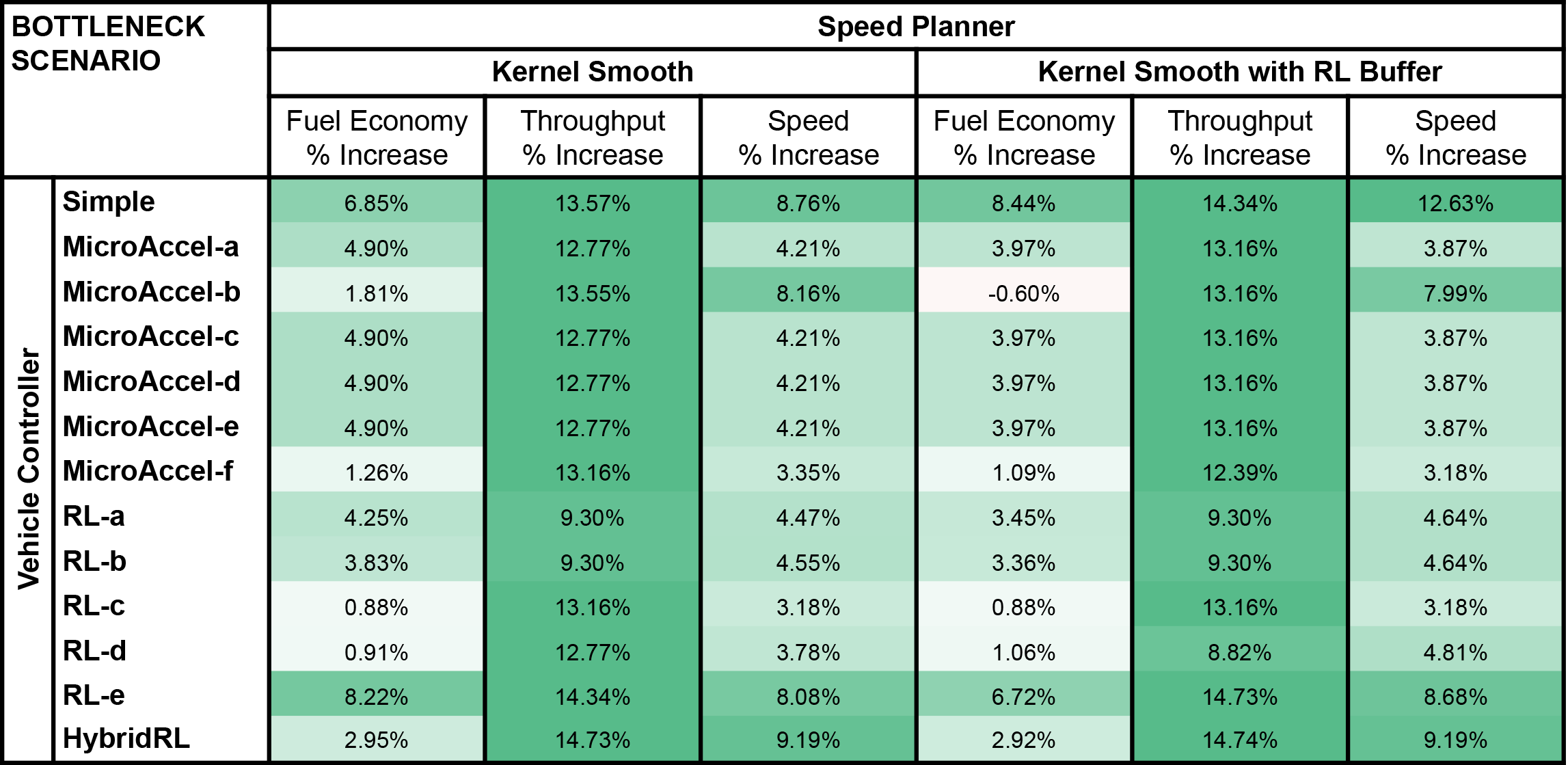}
  \\[\smallskipamount]
  \includegraphics[width=0.8\columnwidth]{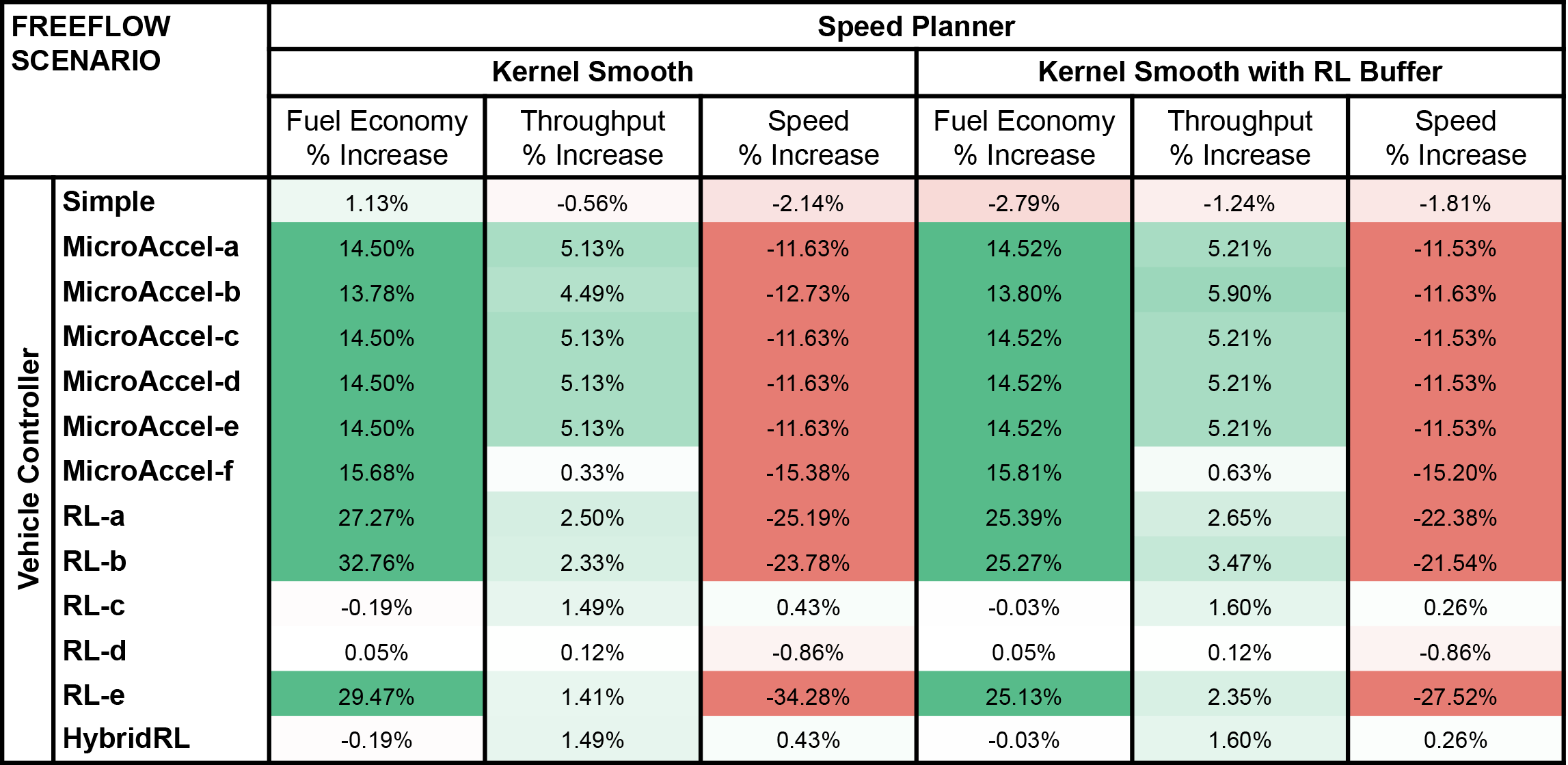}
\caption{Summary of controller assessments prior to MVT deployment. Candidate controller components to the MegaController were simulated under three scenarios (top: shockwave; middle: bottleneck; bottom: freeflow) and evaluated against key performance indicators (fuel economy, throughput, network speed) relative to a baseline simulation of homogeneous human drivers. Speed Planner candidates are shown across the top, and Vehicle Controllers are shown in rows. Key performance indicators are shown in columns. Results are color-coded from improved (green) to neutral (white) to worsened (red).}
\label{fig:freeze_evals}
\end{figure}

\begin{sidebar}{Trajectory Simulator}
\noindent
by Nathan Lichtlé
\section[Trajectory Simulator]{}\label{sbar-TS}
\renewcommand{\thesequation}{S\arabic{sequation}}
\renewcommand{\thestable}{S\arabic{stable}}
\renewcommand{\thesfigure}{S\arabic{sfigure}}

\vspace{-5mm}
\sdbarinitial{E}stablishing a robust, representative, and reliable simulation environment is key for the successful implementation of our wave-smoothing controllers in autonomous vehicles, as it ensures their efficient operation across a wide range of traffic conditions. Our simulator~\cite{icrasim} is based on real human driver data collected on the highway\cite{trajdataseti24}, and has been extensively used for designing, training and evaluating the different types of controls implemented in this work, assessing their safety, energy-reducing performances, robustness and smoothness. This sidebar describes the procedure by which highway trajectory data was gathered and a simulator created from it.

\section{Acquisition of Data}
The trajectory dataset \cite{trajdataseti24} for our study were recorded on a 14.5-kilometer segment of I-24, located southeast of Nashville, Tennessee. An example recorded speed trajectory profile can be seen in Figure~\ref{fig:example_dataset_traj_7050}. An instrumented vehicle is used to gather data which logs vehicle controller area network (CAN) data through \texttt{libpanda}~\cite{bunting2021libpanda} and GPS information from an in-built receiver. The CAN data collection comprises measurements like the speed of the vehicle under consideration (ego vehicle), the relative speed of the lead vehicle (the vehicle in front), the instantaneous acceleration, and the space gap (distance from bumper to bumper).

\sdbarfig{\includegraphics[width=14.0pc]{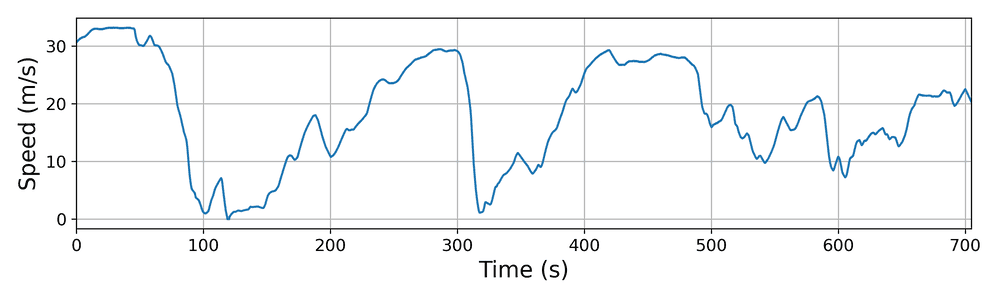}}{Speed vs. time for one of the dataset trajectories, exhibiting large acceleration and breaking patterns that can typically lead to stop-and-go waves.\label{fig:example_dataset_traj_7050}}



\section{Refining Raw Data}
Each drive's raw data is stored in two separate files: a CAN data file and a GPS file. The relevant data are extracted from the CAN file and adjusted to match the GPS time, which is recorded at 10 Hz. The high-frequency CAN data are downsampled and interpolated linearly to align with GPS time, whereas low-frequency CAN data are subjected to linear interpolation to match the 10 Hz GPS time. GPS position data is used to calculate the distance traveled and direction. As westbound data usually demonstrate more consistent congestion, they are primarily used for training, comprising 60 trajectories, which translates to 8.8 hours and 772.3 kilometers of driving.

\section{Analyzing the Dataset}
The trajectory dataset encapsulates a wide variety of traffic conditions ranging from nearly stationary congested traffic to maximum speed free-flow traffic, including diverse acceleration and deceleration patterns associated with stop-and-go traffic. The example trajectory displayed in Figure~\ref{fig:example_dataset_traj_7050} demonstrates the ego vehicle's quick transitions between low and high speeds.

While the primary interest lies in mitigating high-frequency waves that are common in congestion, Figure~\ref{fig:speed_histogram_i24} indicates the tendency of speeds in the training dataset towards the higher end. Despite the possibility of simplifying the learning problem by filtering out high-speed data, congestion zones are often immediately followed by high-speed areas. To ensure our controller's competent behavior at high speeds and during transitions from high to low-speed zones, the training dataset retains both low and high-speed data.

\sdbarfig{\includegraphics[width=14.0pc]{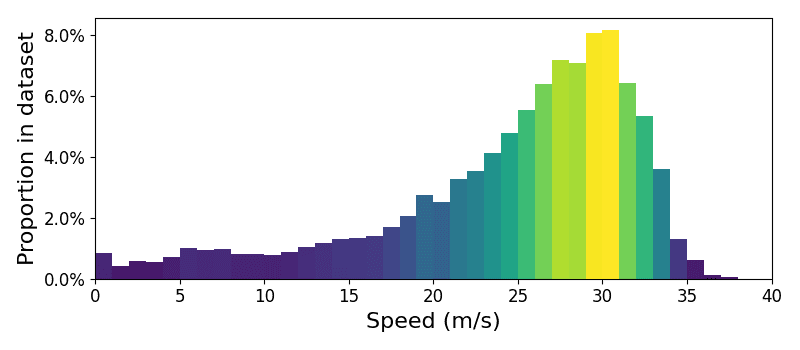}}{Distribution of speeds in the I-24 dataset.\label{fig:speed_histogram_i24}}

\section{Developing the Training and Evaluation Framework}
To exploit the gathered data, a single-lane training environment is designed where an AV follows the trajectory data recorded from human drivers. The human driver is simulated at the front of a vehicle platoon, followed by the AV, and then a number of vehicles operating according to the \emph{Intelligent Driver Model} (see ``\nameref{sbar-CFM}''). This setup guarantees the growth of waves in congestion, due to the string-unstable nature of the chosen IDM parameters. While a comprehensive micro-simulation of I-24 might allow training with more complex vehicle interactions, the proposed simulator focuses on realistic driving dynamics representative of the highway's wave types and drivers' reactions to wave formation. The simulator's efficiency is also commendable as it achieves 2000 steps per second, whereas a micro-simulation of the complete 14-kilometer section would be computationally expensive due to the thousands of vehicles in congestion.

\end{sidebar}

\begin{sidebar}{Car-Following Models}
\noindent
by Nour Khoudari, Benjamin Seibold
\section[Car-Following Models]{}\label{sbar-CFM}
\vspace{-5mm}
\renewcommand{\thesequation}{S\arabic{sequation}}
\renewcommand{\thestable}{S\arabic{stable}}
\renewcommand{\thesfigure}{S\arabic{sfigure}}

\sdbarinitial{C}\emph{ar-following models} (CFMs)~\cite{brackstone1999car,garavello2016models} are systems of ordinary differential equations (ODEs) describing the dynamics of each vehicle on the road, where drivers react to the changes in the relative positions of the vehicle ahead. The ODEs could describe the vehicle velocity only and those are first-order models, or the velocity and the acceleration and those are second-order models. Second-order CFMs are of the form
\begin{equation}
\dot{v}(t) = f(s(t), v(t), \Delta v(t))\;,
\end{equation}
where $s$ is the gap to the vehicle ahead measured in meters, $v$ is the velocity of the vehicle, commonly measured in $\text{m}/\text{s}$, and $\Delta v$ is the velocity difference or the approaching rate to the vehicle ahead, commonly measured in $\text{m}/\text{s}$.
An elegant CFM is the Optimal Velocity Model (OVM)~\cite{bando1995dynamical}
\begin{equation}
f(s,v,\Delta v)_{\text{OVM}} =\alpha\left[V(s)- v\right]\;,
\end{equation}
where the optimal velocity function, $V(s)$, determined by the gap $s$ to the vehicle ahead, is a positive monotonically increasing function, with $s\to\infty$ asymptote at the speed limit. Some variations of this model were proposed to avoid car collisions, an example is the Optimal Velocity Follow the Leader Model (OVM-FtL):
\begin{equation}
\label{eqn: OV-FtL}
    f(s,v,\Delta v)_{\text{OVM-FtL}} = \alpha\left[V(s)- v\right]+ \beta\left
    [\frac{\Delta v}{s^\nu}\right]\;,
\end{equation}
where $\nu$ is a positive exponent, and $\beta$ is a positive braking coefficient (measured in $\text{m}^\nu/\text{s}$).
Another example of second-order CFMs is the Intelligent Driver Model (IDM), introduced in~\cite{treiber2000congested} and suitably adapted and used in this work. The IDM acceleration function is
\begin{equation}
f(s,v,\Delta v)_{\text{IDM}} =a\left[1-\left(\frac{v}{v_{0}}\right)^{\delta}-\left(\frac{s^{*}(v,\Delta v)}{s}\right)^{2}\right]\;,
\end{equation}
where $s^{*}(v,\Delta v) = s_{0}+vT+\frac{\max\{0,v\Delta v\}}{2\sqrt{ab}}$. Here $v_0$ is the desired velocity on an empty road (measured in $\text{m}/\text{s}$), $s_0$ represents the minimum spacing between vehicles (measured in $\text{m}$), $T$ is the minimum possible time gap to reach the vehicle ahead (measured in $\text{s}$), $\delta$ is an acceleration exponent, and $a$ and $b$ are the maximum vehicle acceleration and minimum desired comfortable deceleration, respectively (measured in $\text{m}/\text{s}^2$).
\end{sidebar}

\section{Open Road Field Operational Test}
Following the controller selection, this section introduces the implementation details of the selected controllers, experimental design for the 100 AV deployment, and contemporaneous data collection procedures.



\subsection{Server-side implementation}
As indicated in Figure \ref{fig:planner}, the server-side implementation includes the construction of the database, the API, and the Speed Planner algorithm scripts. The database schema is designed following the star schema data model, consisting of dimension tables (for storing static metadata) and fact tables (for storing quantitative data). The tables include:
\begin{itemize}
    \item \texttt{dim\_vehicle} for storing the vehicle metadata, route and lane assignment;
    \item \texttt{dim\_inrix\_segment} for storing INRIX road segment data for I-24;
    \item \texttt{dim\_i24\_segment} for storing finer-grained road segment data, which supports TSE and Speed Plan profiles after data fusion;
    \item \texttt{fact\_vehicle\_ping} for storing the realtime AV ping information;
    \item \texttt{fact\_vehicle\_observation} for storing extracted vehicle observations from the AV ping table;
    \item \texttt{fact\_inrix\_estimate} for storing realtime INRIX data;
    \item \texttt{fact\_speed\_planner} for storing Speed Plan profiles.
\end{itemize}
The \texttt{fact\_vehicle\_ping} and \texttt{fact\_inrix\_estimate} tables are continually being populated with realtime updates at 1Hz. After processing these input data via the Python-based Speed Planner algorithm, the Speed Plan is written to the \texttt{fact\_speed\_planner} table, which is exposed to the internet via a PHP-based HTTP API.

\subsection{Vehicle-side implementation}
\begin{figure}[h!]
    \centering
    \includegraphics[width=\columnwidth]{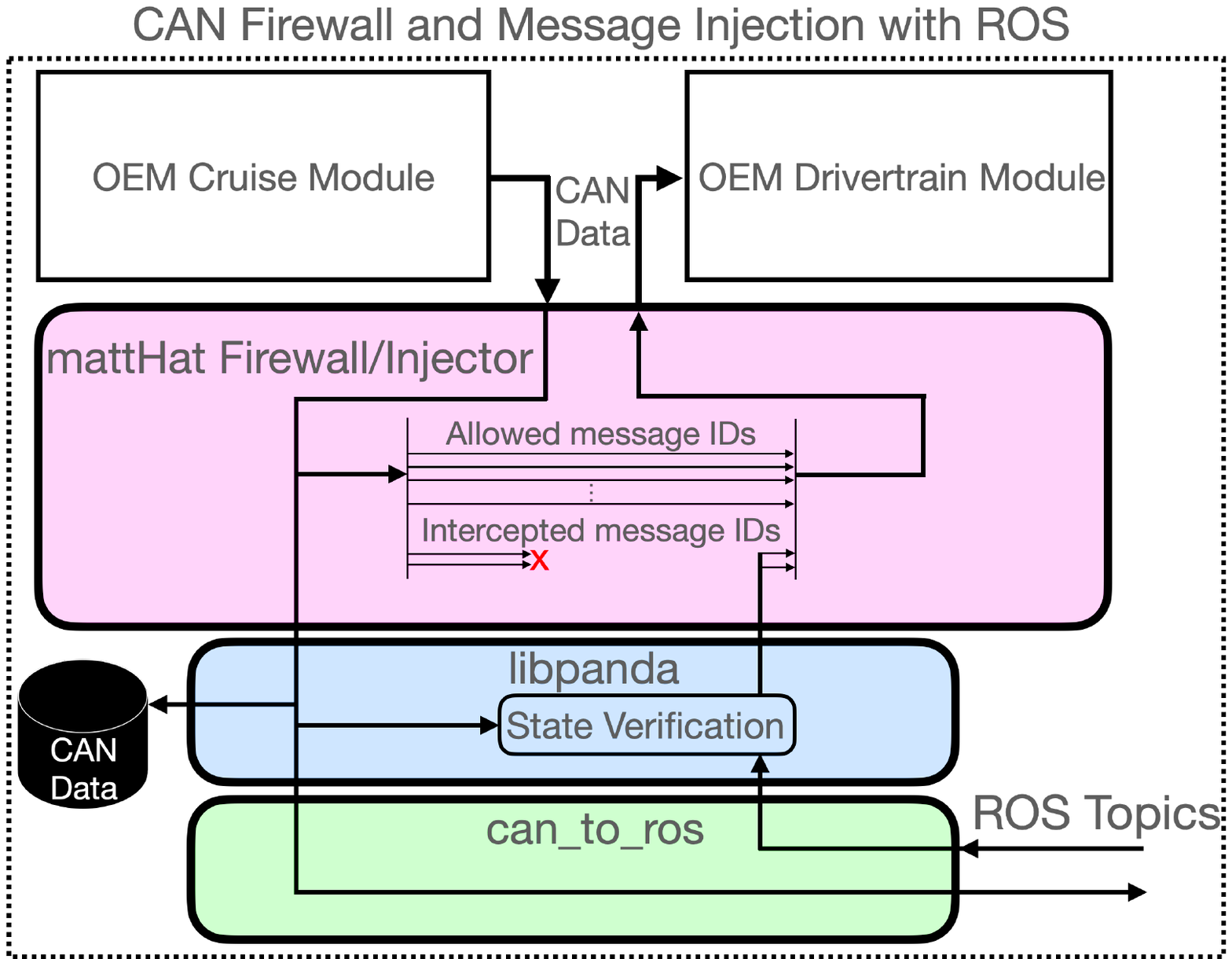}
    \caption{ A special CAN interface called the mattHat firewalls messages between OEM modules at the hardware level, and can replace blocked messages with third party messages from software with libpanda.  Libpanda provides direct CAN recording as well as verification of CAN data to be sent by checking OEM's CAN states.  CAN messages can be read and sent through ROS topics provided by adapters in can\_to\_ros. }
    \label{fig:caninjection}
\end{figure}
The vehicle-side implementation includes several software libraries and a mix of custom and off the shelf hardware components. 

In the last 30 years, vehicles have been gradually increasing the implementation of the open \acronym{Controller Area Network} (CAN) protocol for in-vehicle networking communication between various electrical modules.  The usage of the CAN bus lets automotive manufacturers use minimal wiring to design complex systems involving anything from engine diagnostics, infotainment, security, emissions, and (in more recent years) \acronym{adaptive cruise control} (ACC) and \acronym{Lane Keep Assist} (LKA). The ACC system on cars typically involves a sensor module to measure the leading vehicle dynamics and a separate controller module, which communicates with the vehicle's transmission and engine. Since these modules are physically located in different parts of the vehicle they communicate using CAN busses, where listening CAN analyzers can record the information.  Figure~\ref{fig:caninjection} shows the method of tapping into a CAN bus for message reading.


\begin{sidebar}{Vehicle Interfacing}
\noindent
by Matthew Bunting
\section[Vehicle Interfacing]{}\label{sbar-HW}
\vspace{-5mm}

\renewcommand{\thesequation}{S\arabic{sequation}}
\renewcommand{\thestable}{S\arabic{stable}}
\renewcommand{\thesfigure}{S\arabic{sfigure}}

\sdbarinitial{T}he set of hardware involves both off-the-shelf and custom components assembled as a stack of Printed Circuit Boards (PCB).  Figure~\ref{sfig_hardware_stack} shows the relationship of hardware components installed in each car.

\section{Hardware}
A Raspberry Pi 4 served as the main computer to run controllers.  This was chosen due to its wide open-source support, and its availability and low cost.  The operating system was 64-bit Raspbian Lite.  We chose Raspbian instead of other Linux distributions since it was the only actively fully supported version from the hardware manufacturer.

Attached to the Raspberry PI was an x728 battery backup \acronym{Uninterrupted Power Supply} (UPS).  An experiment lifecycle began turning on the car and therefore providing power to the Raspberry Pi, then ended when the vehicle parked and shut off at our headquarters.  The UPS would continue to provide power to the Pi, and signal that the Pi should stop processed of custom control and upload the experiment's data over WiFi.

To provide internet to the Pi during an experiment, needed for receiving control setpoints, a separate mobile hotspot was connected to the Pi's ethernet.  Our chosen hotspot was an industrial grade Cradlepoint IBR900 and IBR600. This hardware setup was shown to be effective for societal-scale experiments~\cite{richardson_analysis}. 

A uBlox M8 series USB GPS module provided the system with positional tracking.  This sensor was also used to synchronize the Raspberry Pi's system clock to GPS time, to ensure that later data processing would involve minimal manual realignment.

The mattHAT served as the interface for the vehicles, with the majority being a 2023 Nissan Rogue.  The only standard interface on vehicles is the \emph{Onboard Diagnostics} (OBD) port, however this only provides useful-yet-minimal data like the \emph{Vehicle Identification Number} (VIN)  While modern vehicles have a rich set of sensors and actuators using standardized CAN bus for communication, they are minimally documented to prevent third parties from interpreting sensors and from sending actuation commands.  This leads to a significant effort to decode CAN signals.  While off-the-shelf CAN decoders exist, our solution had to be custom in order to send low-level electrical signals to operate the stock ACC unit.  This was possible for the Nissan Rogue by applying an electrical resistance to spoof button presses, requiring a custom circuit.  The mattHAT was designed with these three communication components in mind, to read the VIN over OBD, read vehicle sensors and state information over specific CAN busses, and to send low-level ACC control commands.

\sdbarfig{\label{sfig_hardware_stack}\includegraphics[width=14.0pc]{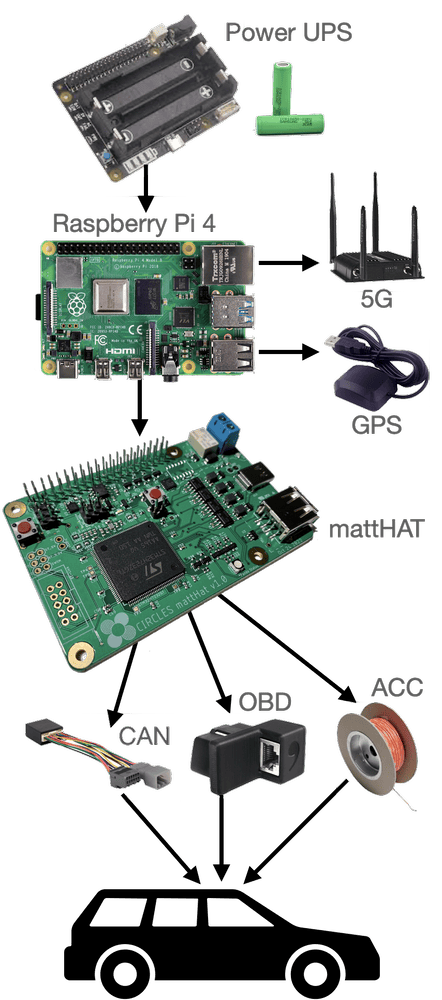}}{The set of hardware installed in each vehicle.}
\section{Software}
\texttt{libpanda} is a low-level C++ library to operate the mattHAT, written with a focus on data timeliness and low CPU usage~\cite{bunting_libpanda}. \texttt{libpanda} abstracted the vehicle interface in the form of an observer design pattern to easy write software for reading CAN and GPS data, and to send control commands wither through CAN message injection or through the mattHAT's ACC interface.  In addition to low-level operation of the vehicle, \texttt{libpanda} also features a set of utilities to manage the Raspberry Pi for functions like automatic data upload, \emph{Over-the-Air} (OTA) updates, and reporting system status.

\end{sidebar}

\begin{sidebar}{\continuesidebar}
\renewcommand{\thesequation}{S\arabic{sequation}}
\renewcommand{\thestable}{S\arabic{stable}}
\renewcommand{\thesfigure}{S\arabic{sfigure}}
Using \texttt{libpanda}, a ROS node was designed to translate the various vehicle functions into the form of ROS topics.  A project named \texttt{can\_to\_ros} provided pre-written middleware so that controller designers could use tools like MATLAB's Simulink to generate code in the form of C++ to control the vehicle~\cite{elmadani_can}.  \texttt{can\_to\_ros} was also expanded to handle a heterogeneous fleet, due to the mixture of models and make of vehicles~\cite{nice_middleware} \cite{bhadani_approaches}.  In conjunction with \texttt{can\_to\_ros} and \texttt{libpanda}, a data analysis tools named \texttt{bagpy}~\cite{bhadani_bagpy} and \texttt{strym}~\cite{bhadani_strym} were used to quickly analyze and decode recorded data.  Data from the signal decoding and system characterization process for a 2020 Toyota RAV4 has also been made publicly available~\cite{bunting_data}.


\end{sidebar}

While automotive manufacturers provide information on the structure of the in-vehicle network in the form of wiring diagrams, specific information being sent on the CAN bus is typically a closely guarded secret, likely due to both trade secrets and safety concerns. However, some companies (such as comma.ai~\cite{comma_ai} and Intrepid Control Systems~\cite{intrepid_control_sys}) are entering the space of custom vehicle autonomy by selling modules that intercept CAN messages between these modules and provide their own inputs based on custom controllers. This is only possible after spending time decoding messages and reverse engineering the protocols needed to inject custom messages.  Compared to basic CAN reading, Figure~\ref{fig:caninjection} shows the different electrical architecture needed for CAN message interception and injection between OEM modules.

\begin{figure}[h!]
    \centering
    \includegraphics[width=\columnwidth]{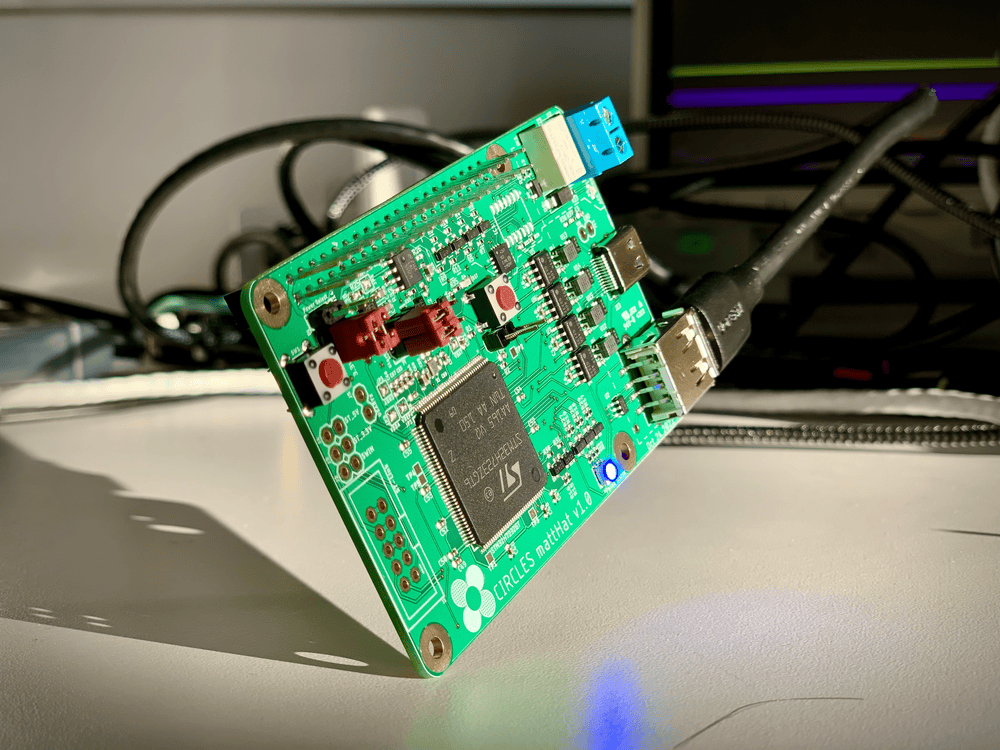}
    \caption{ The mattHat: a custom designed circuit board rapidly manufactured for both CAN interfacing and ACC button command spoofing.  }
    \label{fig:matthat}
\end{figure}

In this work our solution is based on a similar principle, however additional protocols on the 2023 Nissan Rogue (which comprise the vast majority of our control fleet) prevented the injection of commands, though CAN data could still be read for real-time controller inputs.  Due primarily to supply chain issues and additionally to an inability to inject CAN messages, a custom circuit board was developed, coined mattHAT (Hardware Attached on Top).  A mattHat can be seen in Figure~\ref{fig:matthat}.  The mattHAT acted as a CAN interface and also provided methods of sending ACC button press commands using custom circuitry.  The board was designed to plug into a Raspberry Pi 4.  Custom wire harness cables were designed and built for vehicle installation across the 97 Nissan Rogue vehicles, targeting specific CAN busses related to the vehicle's ACC state.  

A C++ based library named \texttt{libpanda}~\cite{bunting2021libpanda} was developed to operate the mattHAT's CAN interface and button spoofer as well as record raw CAN and GPS data.  \texttt{libpanda} also supports a USB-based GPS module to both record the vehicle's position and to synchronize the system's time.  A tool named Strym was also developed to decode and quickly analyze libpanda's recorded data for CAN signal decoding and classification~\cite{bhadani2022strym,ngo2021lightweight,ngo2022canclassify}.  To allow control designers to easily make use of \texttt{libpanda}, the \acronym{Robot Operating System} (ROS) was installed on the Raspberry Pi.  A set of ROS nodes were designed in the software project named \texttt{can\_to\_ros}~\cite{elmadani2021can,nice2023middleware}, which abstracts various sensors and actuators into a set of ROS topics.  A tool named \texttt{bagpy}  was also developed to quickly process and plot recorded ROS data, in the format of bagfiles~\cite{bhadani2023prototyping}.  With the infrastructure of ROS in place, control designers could use modeling software like MATLAB's Simulink to design controllers and generate code to greatly ease controller integration.

\begin{figure}[h!]
    \centering
    \includegraphics[width=19.0pc]{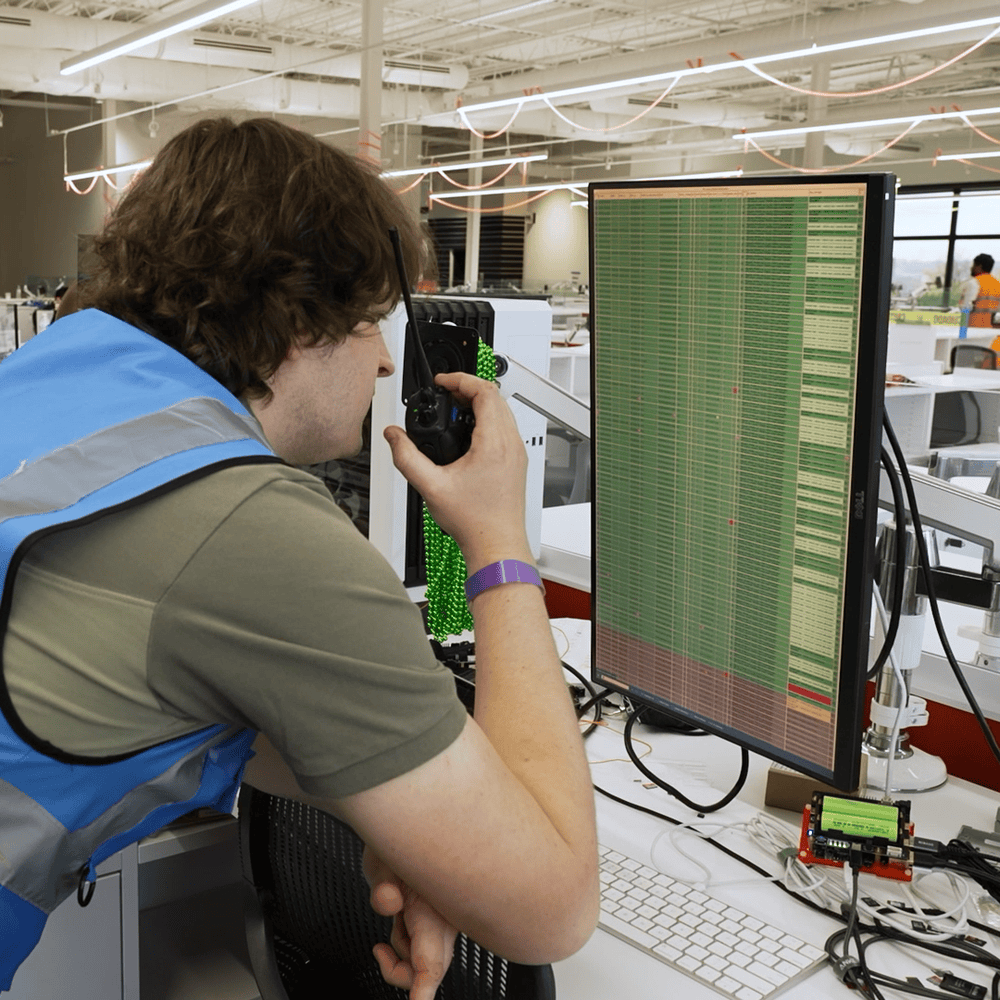}
    \caption{\texttt{piStatus} web interface. Co-author Matthew Bunting monitors the \texttt{piStatus} page to clear vehicles for experiment deployment.}
    \label{fig:pistatus_deployment}
\end{figure}

\texttt{libpanda} also features a set of auxiliary services to manage the vehicle at scale.  A method to automatically perform \acronym{over the air} (OTA) updates was implemented so that vehicles would not need to be manually handled.  An additional server integration named \texttt{piStatus} was built so that the system could regularly post its status to easily assess the system's health.  Figure~\ref{fig:pistatus_deployment} shows headquarters personnel monitoring \texttt{piStatus}, ensuring vehicles pass status checks before deployment.  \texttt{piStatus} would be continuously monitored throughout each experiment to note any hardware issues so that vehicles could undergo maintenance before the next deployment. A shutdown script also worked in tandem with an off-the-shelf battery-backup Pi HAT to perform automatic data uploads at the end of an experimental drive when the vehicle was shut off.  Figure~\ref{fig:antenna} shows a WiFi antenna installed in the center of the vehicle parking lot for granting each vehicle's embedded computer internet access for data upload and OTA updates.  Without these set of services, managing and maintaining a project at this scale would have been impossible.

\begin{figure}[h!]
    \centering
    \includegraphics[width=19.0pc]{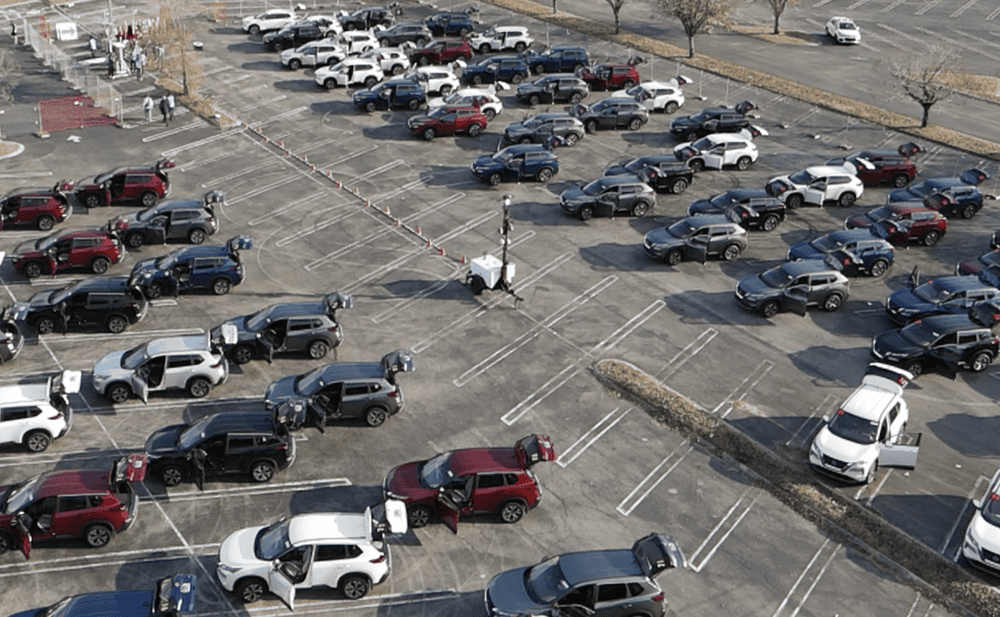}
    \caption{A WiFi antenna is placed in the center of the vehicle parking lot, enabling OTA data uploads and software updates for each Raspberry Pi in each vehicle.}
    \label{fig:antenna}
\end{figure}


\begin{figure*}
    \centering
    \includegraphics[width=\textwidth]{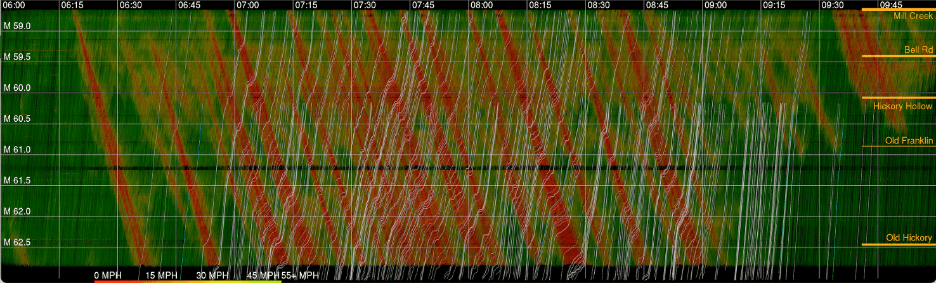}
    \caption{Time (horizontal axis) space (vertical axis) diagram generated by I-24 MOTION~\cite{gloudemans202324} and associated visualization library~\cite{10.1145/3576914.3587710} during the MVT. Vehicle trajectories driving westbound (up) are colored based on the speed of vehicles (green: freeflow to red: congested). Experiment vehicle trajectories are overlaid in white.}
    \label{fig:TS_megacontroller}
\end{figure*}

\subsection{I-24 MOTION}
The CIRCLES team deployed the vehicle controllers on I-24 southeast of Nashville, TN due to the recent creation of a new testbed, known as I-24 \acronym{Mobility Technology Interstate Observation Network} (MOTION)~\cite{gloudemans202324,gloudemans2020interstate}. I-24 MOTION is a four-mile section of I-24 designed to produce ultra-high resolution trajectory data of all vehicles on the roadway for the purposes of traffic science and experimentation on automated vehicles and traffic management. The system produces trajectory data ((Figure~\ref{fig:TS_megacontroller}) using 276 cameras on fixed roadside poles between 110--135ft tall, to minimize visual occlusion. The debut of the I-24 MOTION testbed coincided with the live CIRCLES experiment, for which the testbed is uniquely suited to gather data on the traffic stream impacts of a fleet of test vehicles.

Processing of raw video into vehicle trajectories by I-24 MOTION happens in two stages. 1) A computer vision pipeline~\cite{gloudemans2021vehicle,gloudemans2020interstate} makes the initial vehicle detection and type classification from the raw video~\cite{gloudemans2021vehicle}, including vehicle dimensions by using 3-D bounding boxes. Vehicle classes used by the object detection are: sedan, midsize,  pickup, van, semi, truck, motorcycle. It tracks detected vehicles across adjacent camera views in the vehicle's direction of travel. The computer vision processing is distributed across ten servers with contiguous groups of cameras allocated to each. Vehicles are not tracked between servers, so a vehicle trajectory covering the length of the testbed consists of at least ten fragments; additional fragmentation may occur due to unavoidable occlusion (for example, by overpasses). The computer vision pipeline also converts image space coordinates (where vehicle detections occur) into a roadway coordinate system using a homography transformation calibrated every hour to each camera. 2) Post-processing algorithms~\cite{wang2022automatic} first stitch fragmented trajectories together using an online minimum cost network flow graph problem. Each trajectory is then subject to a reconciliation procedure to ensure feasible and smooth higher order dynamics (acceleration and jerk), formulated as a quadratic program.

Processing of raw video into vehicle trajectories by I-24 MOTION happens in two stages. 1) A computer vision pipeline~\cite{gloudemans2021vehicle,gloudemans2023interstate,gloudemans2023so} makes the initial vehicle detection and type classification from the raw video, including vehicle dimensions by using 3-D bounding boxes. Vehicle classes used by the object detection are: sedan, midsize,  pickup, van, semi, truck, motorcycle. It tracks detected vehicles across adjacent camera views in the vehicle's direction of travel. The computer vision processing is distributed across ten servers with contiguous groups of cameras allocated to each. Vehicles are not tracked between servers, so a vehicle trajectory covering the length of the testbed consists of at least ten fragments; additional fragmentation may occur due to unavoidable occlusion (for example, by overpasses). The computer vision pipeline also converts image space coordinates (where vehicle detections occur) into a roadway coordinate system using a homography transformation calibrated every hour to each camera. 2) Post-processing algorithms first stitch fragmented trajectories together using an online minimum cost network flow graph problem~\cite{wang2023onlinemcf}. Each trajectory is then subject to a reconciliation procedure~\cite{wang2022automatic} to ensure feasible and smooth higher order dynamics (acceleration and jerk), formulated as a quadratic program. A data visualization library~\cite{10.1145/3576914.3587710} assists with interpreting the datasets generated by the system.

Due to the critical nature of data from the MVT, the I-24 MOTION system retained a secure backup of the experiment data and baseline traffic conditions. Raw I-24 MOTION imagery is not accessible outside of the I-24 MOTION administrators, per the testbed's privacy policy. However, this backup allows re-processing of the video and/or the raw vehicle detections from the computer vision pipeline as data generation pipeline advances. The current released version of the data has gone through re-processing rounds to address known errors and limitations, including using hourly re-calibrated homography transformations for camera perspective~\cite{gloudemans2023so}.

\subsection{Experimental Design}
The live traffic experiment took place on I-24 in November 2022. With 100 vehicles to deploy, over 150 drivers were hired from local colleges, security guards, delivery drivers, team members' relatives, and elsewhere. The drivers were trained to adhere to specific assigned routes and lanes and to activate our custom ACC system whenever driving in their assigned lanes. Safety features were implemented such that the controller would default to stock ACC behavior if engaged off of the highway.

The specific routes driven by the AVs were chosen to maximize the penetration rate where the I-24 MOTION system~\cite{gloudemans202324} would capture the effect on the surrounding commuters (bulk traffic). Initially, a single-loop route that would have the AVs circulating from 7:30am to 9:30am was considered. It was determined that this would likely cause increased congestion on the Exit 57 off-ramp, leading to extended queuing and potential spillback onto I-24. For this reason, we divided our AVs into two groups with the two different routes partially overlapped on I-24.


The AVs are released at 6:00am in order for all 100 to be deployed by the target time 7:30am. The drivers are instructed to repeat their driving routes for 2 hours, or sooner if they wanted a break. On a return trip, they exit the highway at Exit 60 (Hickory Hollow), to return to the Field Headquarters parking lot and returned their car keys at a desk before leaving the lot. As AVs return to the lot for a break, other drivers (already on break) head down to a queue to come out to the lot when needed. With this smoothly running rotation system, we were able to keep the 100 AVs on the routes during peak hours. For more details about the routes, and logistical choices made for the experiment, driver training, the daily schedule, or the penetration rate estimates, see the article~\cite{AmeliLivetraffic}.

\begin{figure*}
    \centering
    \includegraphics[width=.99\textwidth]{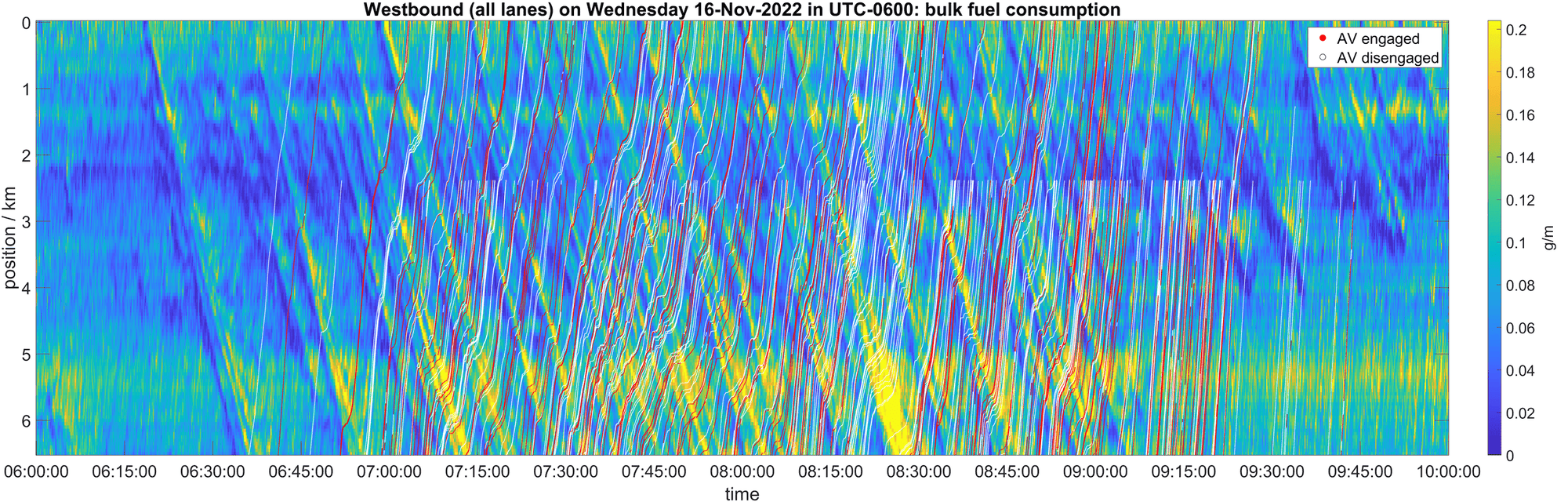}\\
    \includegraphics[width=.99\textwidth]{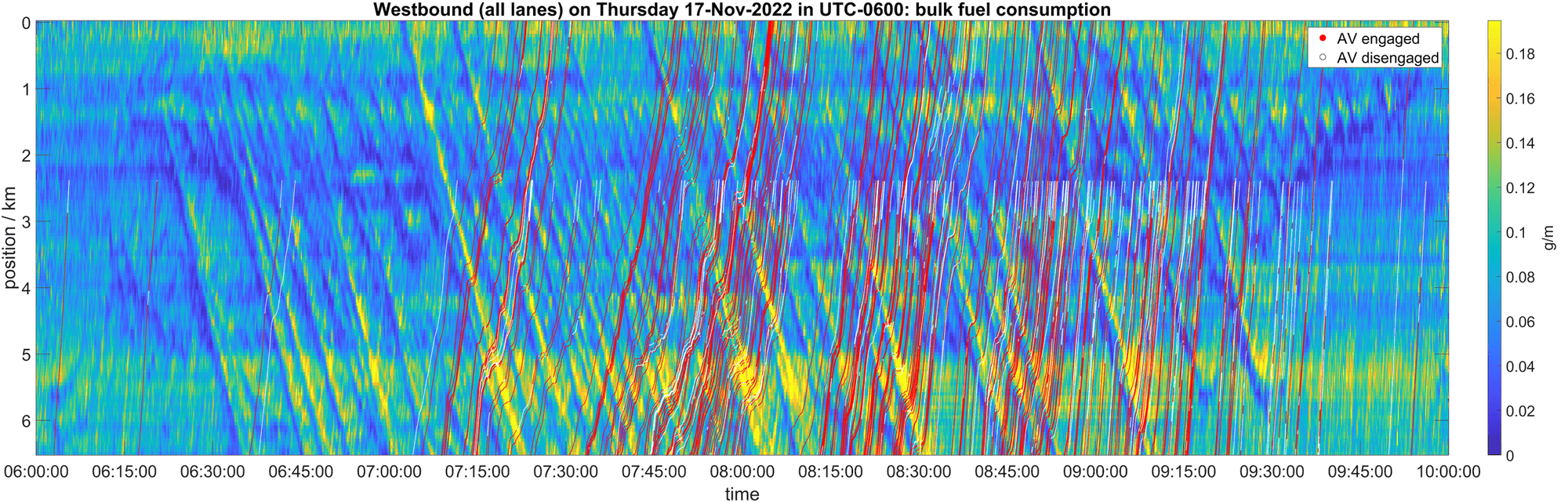}
    \caption{Bulk fuel consumption heatmap in time (horizontal axis) and space (vertical axis), based on I-24 MOTION~\cite{gloudemans202324}  of all vehicles (aggregated over all lanes) driving in the Westbound direction. AV trajectories are overlaid (white: controller not engaged; red: controller engaged). Shown are November 16, 2022 (top) and November 17, 2022 (bottom).}
    \label{fig:TS_bulkfuel}
\end{figure*}

\subsection{Data Release}

Before this publication goes to press, we will have released to the public a dataset from the MVT. This data release will include data collected by the AVs as well as the I-24 MOTION~\cite{gloudemans2020interstate} system. See also~\cite{gloudemans2021vehicle,gloudemans2023so,gloudemans2023interstate} for other datasets generated by I-24 MOTION. Concurrent with this publication’s release, we will issue a companion that demonstrates how to extract the data in order to generate the plots in this paper. Further information on the structure of the data will be provided in the data documentation released with the data.

\section{Results}

The experimental, observational, data collection, and data processing framework described above generated a large amount of data, capturing every single vehicle on a highway during the deployment of 100 controlled vehicles. 
Analogous to previous seminal traffic data sets, such as NGSIM \cite{TrafficNGSIM}, the new data are expected to inspire and enable many subsequent findings. To highlight this potential, we here present some key first findings and insights, based on an analysis of the data  with a macroscopic perspective.

We construct macroscopically meaningful fields in time--space, most prominently a field that shows the energy (in)efficiency of traffic at large on the I-24 highway segment. This is achieved by applying Edie's method~\cite{edie1961car} on boxes of size $h_t \times h_x$, where $h_t = 10\text{s}$ and $h_x = 200\text{m}$, to the I-24 MOTION trajectories to construct the following fields:
\begin{itemize}
\item vehicle density $\rho(t,x)$, as the total vehicle time spent in each box, divided by the size of the box, $h_t\cdot h_x$;
\item flow rate $q(t,x)$, as the total distance traveled in each box, divided by  $h_t\cdot h_x$; and
\item fuel rate density $f(t,x)$, as the total fuel consumed in each box, divided by $h_t\cdot h_x$.
\end{itemize}
From these fields, other meaningful fields are obtained, such as the bulk velocity field $u(t,x) = q(t,x)/\rho(t,x)$, the bulk fuel rate $\phi(t,x) = f(t,x)/\rho(t,x)$, and the bulk fuel consumption $\psi(t,x) = f(t,x)/q(t,x)$. The latter quantity $\psi(t,x)$, measureable for instance in grams per meter, represents the fuel demand per distance traveled of all vehicles in the $h_t \times h_x$ vicinity of the position $(t,x)$. Figure~\ref{fig:TS_bulkfuel} shows $\psi(t,x)$ for two experimental days: Wednesday November 16 and Thursday November 17, 2022. Each plot is overlaid with the trajectories of all control vehicles. This represents the first time that a complete time--space diagram of the energy inefficiency of traffic, based on accurate trajectories of \emph{all} vehicles on the roadway has been provided.

In the same spirit as the purely microscopic Figure~\ref{fig:TS_megacontroller}, one can, for both days shown in Figure~\ref{fig:TS_bulkfuel}, clearly see the traffic waves traveling backwards along the highway, as well as the increased fuel consumption incurred in these waves. The figure also shows the increased fuel demand in the uphill segment between $x = 5\text{km}$ and $x = 6\text{km}$, and the reduced fuel demand in the downhill segment thereafter. The AV trajectories are colored red when the automated controller was activated, and white when the vehicle was under human control. The two shown test days were quite different in terms of the engagement rates of the controllers: on 11/16, the controllers were engaged 38\% of all times, while on 11/17, the engagement rate went up to 78\%. This difference was caused by a combination of increased driver comfort with the automation and a more reliable communication of traffic information to the vehicles on 11/17.

Given the low penetration rate of the control vehicles on the highway, it was not expected that they would completely smooth out all traffic waves---and the plots in Figure~\ref{fig:TS_bulkfuel} confirm that expectation. However, the AVs may still have had some positive contribution on the energy efficiency of the flow at large. Whether that was in fact the case, we first note that the macroscopic plots like in Figure~\ref{fig:TS_bulkfuel} allows for a targeted inspection of different regions of interest in time-space. For instance, on 11/16, there is a distinct region of high fuel inefficiency, around time 08:25:00 and location 5.0km--6.5km; and notably, this high-fuel region coincides with all AV controller being inactive. In contrast, on 11/17 such clusters of inactive AVs did not occur---and the fuel consumption map does not exhibit similarly large high-fuel regions. Another notable anecdote occurs on 11/16 in the wave that goes through $x=2.3\text{km}$ at time 07:00:00. First two active AVs notably dampen the wave; then the wave keeps on growing while four inactive AVs run through it (at $x=3\text{km}$); followed by several active AVs (around $x=4\text{km}$) notably dampening the wave again.

\section{Conclusion}
This work describes the control architectures and implementations of a 100 automated vehicle deployment to improve traffic efficiency on a freeway using a small fraction of automated vehicles.  It is the largest field experiment to use CAVs to regulate the overall traffic flow, and the deployment strategy enabled algorithms from diverse fields spanning model based control to reinforcement learning. The control strategy presented in this work was a hierarchical control approach in which the upper level speed planner provided target velocities to a lower level control law responsible for performance and safety.  These algorithms were  deployed in  the largest field experiment of its kind, on a heterogeneous vehicle fleet using low cost vehicular instrumentation. The data from the vehicles were combined with datasets generated from I-24 MOTION~\cite{gloudemans202324,gloudemans2020interstate}, providing a large data resource for further study on the interaction of control vehicles on bulk traffic flow.

\section{Acknowledgement}

This material is based upon work supported by the National Science Foundation under Grants CNS-1837244 (A.~Bayen), CNS-1837652 (D.~Work), CNS-1837481 (B.~Piccoli), CNS-1446715 (B.~Piccoli), CNS-1446690 (B.~Seibold), CNS-1446435 (J.~Sprinkle), CNS-1446702 (D.~Work), CNS-2135579 (D.~Work, A.~Bayen, J.~Sprinkle, J.~Lee), and by the French CNRS under the grant IEA SHYSTRA and PEPS JCJC (A.~Hayat). This material is based upon work supported by the U.S.\ Department of Energy’s Office of Energy Efficiency and Renewable Energy (EERE) under the Vehicle Technologies Office award number CID DE--EE0008872. The views expressed herein do not necessarily represent the views of the U.S.\ Department of Energy or the United States Government.  The authors are grateful for the additional support provided by C3AI, Amazon AWS, Siemens, Toyota, GM, Nissan, Caltrans, CCTA, KACST, and the Tennessee Department of Transportation.

\section{Authors}
The CIRCLES Consortium~(Figure~18) 
 is led by: \begin{IEEEbiography}{Jonathan W.~Lee }{\,} (jonny5@berkeley.edu) Jonathan Lee received the B.S.~degree in engineering physics from the University of California, Berkeley and M.S.~and Ph.D.~degrees in mechanical engineering from Rice University. At Sandia National Laboratories (2011--13), he completed his postdoctoral appointment studying electrical and material properties via molecular dynamics simulations. He subsequently served as a senior data scientist and product manager on various teams at Uber Technologies, Inc.~(2014-2019). Since 2019, he has served as an engineering manager at the University of California, Berkeley and the program manager and Chief Engineer of CIRCLES.\end{IEEEbiography}    \begin{IEEEbiography}{Daniel B.~Work }{\,} (dan.work@vanderbilt.edu) Daniel Work is a Professor of Civil and Environmental Engineering and the Institute for Software Integrated Systems, at Vanderbilt University. His interests are in transportation cyber-physical systems. \end{IEEEbiography}  \begin{IEEEbiography}{Benjamin Seibold }{\,} (seibold@temple.edu) Benjamin Seibold is a Professor of Mathematics and Physics, and the Director of the Center for Computational Mathematics and Modeling, at Temple University. His research areas, funded by NSF, DOE, DAC, USACE, USDA, and PDA, are computational mathematics (high-order methods for differential equations, CFD, molecular dynamics) and applied mathematics and modeling (traffic flow, invasive species, many-agent systems, radiative transfer). \end{IEEEbiography}  \begin{IEEEbiography}{Jonathan Sprinkle }{\,} (jonathan.sprinkle@vanderbilt.edu) Jonathan Sprinkle is a Professor of Computer Science at Vanderbilt University since 2021. Prior to joining Vanderbilt he was the Litton Industries John M. Leonis Distinguished Associate Professor of Electrical and Computer Engineering at the University of Arizona, and the Interim Director of the Transportation Research Institute. From 2017-2019 he served as a Program Director in Cyber-Physical Systems and Smart \& Connected Communities at the National Science Foundation in the CISE Directorate.  \end{IEEEbiography}  \begin{IEEEbiography}{Benedetto Piccoli }{\,} (piccoli@camden.rutgers.edu) Benedetto Piccoli is University Professor and
the Joseph and Loretta Lopez Chair Professor
of Mathematics at Rutgers University -
Camden. He also served as Vice Chancellor for
Research.
His research interests span various areas of
applied mathematics, including control theory,
traffic flow on networks, crowd dynamics, math
finance and application to autonomous driving,
population health and bio-medical systems. He
is author of more than 300 research papers and
7 books and is the founding editor of Networks
and Heterogeneous Media.
Piccoli is the recipient of the 2009 Fubini Prize,
Plenary speaker at ICIAM 2011, and 2012
inaugural Fellow of American Mathematical
Society. \end{IEEEbiography}  \begin{IEEEbiography}{Maria Laura Delle Monache }{\,} (mldellemonache@berkeley.edu) Maria Laura Delle Monache is an Assistant Professor in the Department of Civil and Environmental Engineering at the University of California, Berkeley. Dr. Delle Monache's research lies at the intersection of transportation engineering, mathematics, and control and focuses on modeling and control of mixed autonomy large-scale traffic systems. \end{IEEEbiography}  \begin{IEEEbiography}{Alexandre M.~Bayen }{\,} (bayen@berkeley.edu) Alexandre Bayen is the Associate Provost for Moffett Field Program Development at UC Berkeley, and the Liao-Cho Professor of Engineering at UC Berkeley. He is a Professor of Electrical Engineering and Computer Science and of Civil and Environmental Engineering (courtesy). He is a visiting Professor at Google. He is also a Faculty Scientist in Mechanical Engineering, at the Lawrence Berkeley National Laboratory (LBNL). From 2014 - 2021, he served as the Director of the Institute of Transportation Studies at UC Berkeley (ITS). \end{IEEEbiography}    

\begin{figure*}[ht!]
\centerline{\includegraphics[width=40.0pc]{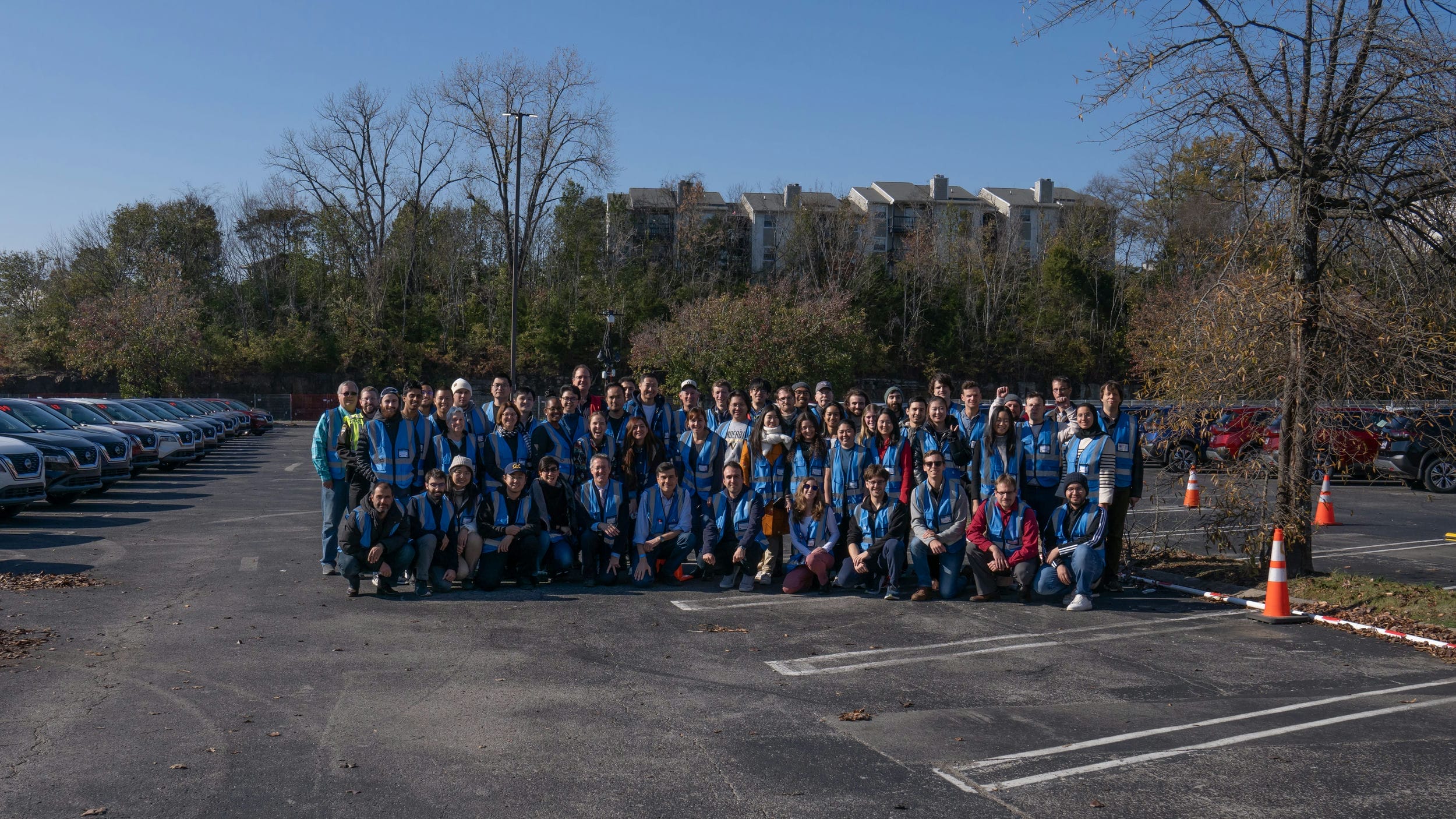}}
\label{fig:teamPhoto}
\caption{The authors during the 100 vehicle experiment in which the MegaController was run.}
\end{figure*}

\bibliographystyle{IEEEtran}
\bibliography{references, csm_v3, CIRCLES-key-papers}

\endarticle

\end{document}